\documentclass[sigconf,nonacm]{acmart}
\AtBeginDocument{%
  }

\usepackage{enumitem}
\usepackage{subcaption}

\newcommand{\popular}{\textsc{r/popular}}
\newcommand{\numparticipants}{\textcolor{black}{600}}

\newcommand{\togetherhdi}[3]{($#1$, 95\% HDI [$#2$, $#3$])}
\newcommand{\separatehdi}[3]{$#1$ (95\% HDI [$#2$, $#3$])}
\newcommand{\onlyhdi}[2]{(95\% HDI [$#1$, $#2$])}
\newcommand{\counts}[2]{($#1\%$, $n=#2$)}
\newcommand{\n}[1]{($n=#1$)}

\begin{document}

% \title{The Ranking Effect: How Algorithmic Rank Shapes User Attention on Social Media}
% \title{The Ranking Effect: How Algorithmic Rank Influences User Attention on Social Media}
\title{The Ranking Effect: How Algorithmic Rank Influences Attention on Social Media}

\author{Jackie Chan}
\email{jackiec3@illinois.edu}
\affiliation{%
    \institution{University of Illinois Urbana-Champaign}
    \city{Urbana}
    \state{Illinois}
    \country{USA}
}

\author{Fred Choi}
\email{fc20@illinois.edu}
\affiliation{%
    \institution{University of Illinois Urbana-Champaign}
    \city{Urbana}
    \state{Illinois}
    \country{USA}
}

\author{Koustuv Saha}
\email{ksaha2@illinois.edu}
\affiliation{%
    \institution{University of Illinois Urbana-Champaign}
    \city{Urbana}
    \state{Illinois}
    \country{USA}
}

\author{Eshwar Chandrasekharan}
\email{eshwar@illinois.edu}
\affiliation{%
    \institution{University of Illinois Urbana-Champaign}
    \city{Urbana}
    \state{Illinois}
    \country{USA}
}


\begin{abstract}
    Social media feeds have become central to the Internet. Among the most visible are \textit{trending feeds}, which rank content deemed timely and relevant. To examine how feed signals influence behaviors and perceptions, we conducted a randomized experiment \n{585} simulating Reddit's \popular{} feed. By having participants view identical sets of posts in different orders, we isolate the effects of rank and social proof on engagement and perceived relevance, trustworthiness, and quality. We found that lower-ranked posts received about 40\% less engagement, despite participants rarely reporting rank as a factor in their choices. In contrast, neither rank nor social proof shifted perceptions across the three dimensions. We also observed demographic patterns: older participants were more skeptical of trending content, while those with less formal education expressed greater trust. Overall, our findings show that algorithmic curation implicitly steers attention, with implications for platform design, research on algorithmic influence, and policy.
\end{abstract}




% \keywords{Do, Not, Us, This, Code, Put, the, Correct, Terms, for,

\maketitle

\section{Introduction}

Social media platforms highlight viral content on ``trending'' or ``popular'' feeds, which function much like the front page of a newspaper by offering a quick glimpse of what is most active, topical, and widely discussed. Although labeled differently across platforms---such as the ``Trending'' tab on X/Twitter, or the ``Explore'' sections on YouTube, Pinterest, and LinkedIn---their purpose is similar: to surface content deemed broadly relevant or engaging at a given moment. On Reddit,
% ---the focus of this study---
the \popular{} feed is the default view for non-logged-in visitors, serving as an entry point into the platform's content ecosystem. These feeds are assembled through algorithmic processes that classify, filter, and \textit{rank} content using engagement signals such as views, clicks, and shares~\cite{eckles-senate-hearing}. While effective at surfacing content, their internal mechanisms remain opaque---both to safeguard proprietary systems and to prevent manipulation---leaving users uncertain about how rankings are determined and researchers unable to fully assess their effects on behavior~\cite{trending-is-trending}.

In particular, there is extensive research on how rankings in search engine result pages (SERPs) influence behaviors and perceptions, including the well-documented search engine manipulation effect (SEME)~\cite{joachims-clicks, biased-search-debated-topics, seme-politics-candidates}. Beyond search, researchers have examined how algorithmic rankings shape outcomes in other domains---such as music markets~\cite{watts-music-markets}, job candidate visibility on LinkedIn~\cite{linkedin-ranking}---and how to design better ranking systems that consider factors beyond just user engagement~\cite{cura, alexandria-reranking, embedding-democratic-values, embedding-values-bernstein}. However, while these studies shed light on the influence of ranking in various contexts, there remains little empirical work examining how ranking on \textit{trending feeds}, and more broadly social media, specifically affects users' perceptions of the content they interact with. Given the substantial traffic trending feeds attract, understanding the potential biases they introduce through ranking is critical.

\subsection{Study Overview}

To address this gap, we conduct a large-scale \n{585} randomized, between-subjects user experiment in which participants browsed systematically-varied screenshots of Reddit's \popular{} feed. The experiment examines how a post's rank and the presence of social proof influence both engagement choices and perceptions of relevance, trustworthiness, and content quality. Following \citet{social-proof-definition}, we define social proof as the tendency of individuals---in this case, social media users---to copy the behaviors of others, operationalized here as post-level engagement metrics (i.e., number of comments and upvotes). All of these dimensions are grounded in prior work on user perceptions~\cite{munmun-relevance, tweeting-believing}, content quality frameworks~\cite{content-quality-frameworks}, and motivations for engaging with trending feeds~\cite{trending-is-trending}. To isolate the effects of rank and social proof, we collect a corpus of \popular{} snapshots and create controlled variations in two ways: (1) rotating the \textit{same} set of posts across different ranks, and (2) either displaying or hiding social proof.

% First, we aim to understand how users browsing these feeds decide which posts to engage with (i.e., click on). Users, when deciding which posts to engage with, may consider a variety of factors, from the community (i.e., subreddit) a post originates from, to the content itself, to the amount of engagement it has received, to more implicit cues such as its rank on the feed. Accordingly, we ask our first research question:

First, we aim to understand how users browsing these feeds decide which posts to engage with (i.e., click on). In making these choices, users may draw on factors such as the originating community (i.e., subreddit), the post's content, visible engagement metrics, or implicit cues like its rank on the feed. Accordingly, we ask our first research question:

\begin{enumerate}[leftmargin=1cm] 
    \item[\textbf{(RQ1)}] To what extent do the following factors influence user engagement decisions on Reddit's trending feed (\popular{}): (a) feed position, (b–c) social proof (number of upvotes and comments), (d) origin subreddit, and (e) post content?
\end{enumerate}

Similarly, we hypothesize that the perceptual dimensions we identified will guide which posts participants choose to engage with:

\begin{enumerate}[leftmargin=1cm]
    \item[\textbf{(H2)}] Relevance, trustworthiness, and content quality significantly influence whether users engage with posts.
\end{enumerate}

Because we collect demographic information from participants, we also examine whether these characteristics affect engagement decisions and perceptions:

\begin{enumerate}[leftmargin=1cm]
    \item[\textbf{(RQ3)}] What effect, if any, do age, gender, and educational attainment have on (a) which posts users choose to engage with on Reddit's trending feed, and on how users perceive a post’s (b) relevance, (c) trustworthiness, and (d) content quality?
\end{enumerate}

Given the diversity of content on the \popular{} feed, we further ask:

\begin{enumerate}[leftmargin=1cm]
    \item[\textbf{(RQ4)}] What effect, if any, does post content (e.g., a post's origin subreddit, content type, or topic) have on the post's (a) likelihood of being engaged with, and on its perceived (b) relevance, (c) trustworthiness, and (d) content quality?
\end{enumerate}

Finally, we hypothesize effects related to the two variables manipulated in our study: rank and social proof (i.e., number of comments and upvotes):

\begin{enumerate}[leftmargin=1cm]
    \item[\textbf{(H5)}] A post placed in a higher position on \popular{} will have a higher likelihood of being (a) engaged with, and will be perceived as having greater (b) relevance, (c) trustworthiness, and (d) content quality, compared to the same post in a lower position.
    \item[\textbf{(H6)}] The presence and degree of social proof (e.g., number of upvotes or comments) is positively associated with a post's (a) likelihood of being engaged with, and perceived (b) relevance, (c) trustworthiness, and (d) content quality.
\end{enumerate}

How we answer and validate these research questions and hypotheses is described in Section~\ref{sec:design}, but these research questions and hypotheses define the focus of our user experiment.

\subsection{Summary of Contributions}

This paper presents a large-scale ($n=585$) randomized, between-subjects user experiment simulating Reddit's \popular{} feed and examining how rank and social proof (i.e., post-level engagement metrics) shape user engagement and perceptions of content along three dimensions: relevance, trustworthiness, and quality.
On algo\-rithmically-ranked feeds like \popular{}, the rank of a post---i.e., the post's position within the feed---cannot be modified in real-time by external researchers. To overcome this challenge, we design a novel approach to systematically create controlled variations in screenshots of \popular{} and examine what happens \textit{when the same post is shown at different ranks on the feed}.

Our study presents several key findings:

\begin{enumerate}
    % \item \textbf{Rank drives engagement but not perceptions.} Posts placed lower in the feed had significantly lower odds ($\approx 40\%$ less) of being selected compared to those at the top, even though participants rarely self-reported rank as a factor. Yet perceptions of relevance, trustworthiness, and quality did not vary significantly across the top 10.
    \item \textbf{Rank drives engagement but not perceptions.} Moving a post from the top of the \popular{} feed to ranks 6--10 lowers its odds of engagement by about $40\%$, even though participants rarely reported rank as influencing their choices. Perceptions of relevance, trustworthiness, and content quality, however, did not differ significantly across the top 10.
    \item \textbf{Demographic differences in trust.} Older participants were more skeptical of posts, rating them as more manipulative than the 25--34 reference group did, while participants without a bachelor’s degree were more trusting of trending content.
    \item \textbf{Limited role of social proof.} Neither the presence nor the degree of engagement metrics (upvotes, comments) significantly influenced engagement or perceptions, suggesting these signals were weak discriminators among the top-ranked posts.
\end{enumerate}

A complete summary of findings can be found in Table~\ref{tab:findings-summary}. Beyond these findings, we situate our results within broader discussions on social media feeds and algorithmic curation. We also consider whether the perceptual dimensions we studied capture the goals participants bring to browsing \popular{}. Overall, this work presents a unique re-ranking experiment on a prominent social media trending feed, highlighting how signals such as rank can meaningfully and subconsciously shift the amount of attention posts receive from end-users.

\section{Related Work}

\subsection{Re-Ranking Social Media Feeds}

Because one of the main factors we manipulate in this paper is the rank of posts on social media feeds---specifically on \popular{}---it is important to situate our work within prior research that has used re-ranking to study online behavior. \citet{reranking-guide} provide a comprehensive guide to conducting field experiments that modify social media feeds in real time. They outline methods for implementing such interventions via browser extensions and describe common stimuli, including up-ranking, down-ranking, and content editing---all of which we incorporate in this study, albeit not in a live setting, as our stimuli are based on previously captured screenshots.

Several studies have explored ranking systems that incorporate factors beyond traditional engagement metrics. For example, \citet{alexandria-reranking} employed LLM-based content classifiers to enable users to re-rank their feeds in real time according to 78 value dimensions, such as wisdom and usefulness. Similarly, \citet{embedding-democratic-values} re-ranked feeds to reduce anti-democratic values and partisan animosity. Both of these works fall under the broader agenda articulated by \citet{embedding-values-bernstein}, which advocates integrating societal values into social media ranking algorithms rather than optimizing solely for engagement. \citet{eckles-senate-hearing} provide a broader overview of research on algorithmic ranking systems on social media.

In contrast to these real-time re-ranking field experiments, our study examines the potential effects that existing rankings and engagement metrics in trending feeds have on user perceptions.

\subsection{Biases on Social Media}

A substantial body of research has documented biases in the ranking of search engine result pages (SERPs)~\cite{ranking-versus-reputation, biased-search-debated-topics, joachims-clicks, serps-conflicting-science}. These studies show that the ranking of a search result can influence how users evaluate its relevance and credibility, independent of the content itself. Given the central role that ranking plays in shaping information exposure, it is natural to ask whether similar effects occur on social media feeds where ranking algorithms determine the order of posts from diverse communities.

One prominent step in this direction is the work of \citet{political-search-bias}, who examined two distinct types of bias in Twitter search results. The first arises from the selection of content shown in the feed, while the second comes from the arrangement of that selected content---a distinction that sets their work apart from other studies. We build on this framework but expand beyond political content to investigate whether ranks in a different type of feed influence how users perceive the same content.

Beyond rank bias, researchers have explored other systematic influences on user perception in social media. For example, profile pictures and usernames on Twitter can affect the perceived credibility of identical claims~\cite{tweeting-believing}. In addition, whether a user opts for a curated (``For You'') feed or a reverse-chronological feed from friends and followers can significantly change the content they encounter~\cite{twitter-sock-puppet-bias, diakopoulos-more-accounts}. User-generated signals can also sway perception: \citet{conformity-trust} found that users often conform to others' opinions of news articles based on the tone of the comment section. Although many studies emphasize that such biases are negative, positive signals---such as creator hearts indicating endorsement by the content creator on YouTube~\cite{creator-hearts-fred}---can shift users' perceptions in a favorable direction. This aligns with other work showing how positive reinforcement can be leveraged in community moderation~\cite{charlotte-quasi}. We extend this line of research by examining the role of social proof---the tendency to look to others to determine appropriate behavior~\cite{social-proof-definition}---in shaping perceptions. In our context, this refers to visible engagement metrics such as the number of comments and upvotes on Reddit posts. Analogous effects have been documented in other domains, such as Airbnb, where the number of reviews and average star ratings influence perceived trustworthiness~\cite{more-stars-reviews}.

In summary, our work investigates whether, and to what extent, both rank and social engagement metrics bias users' perceptions of relevance, trustworthiness, and content quality on a trending feed featuring diverse content from multiple communities. Whereas prior work has largely focused on politics or search queries, we broaden the scope to a multi-topic, multi-community setting.

\subsection{Trends \& Trending Feeds}

In this paper, we use Reddit's \popular{} as the experimental context to examine the impact of rank on user attention. Prior work on Reddit's \popular{} has examined how communities were affected by the attention generated by \popular{}~\cite{chan-community-resilience, chan2022community} and conducted an empirical audit of its ranking algorithm~\cite{chan-popular-audit}.
% While those studies were primarily observational, we build on their work by directly measuring user perceptions through a randomized experiment, testing whether rank and social proof influence how posts are perceived and the likelihood of a user engaging with a post.
While those studies were primarily observational, we build on their work by directly measuring user perceptions through a randomized experiment. In particular, we test whether rank and social proof influence how posts are perceived and whether they affect the likelihood of user engagement.

Related research has examined trending mechanisms on other platforms. For instance, \citet{coordinated-campaigns} studied ``trending topics'' on Twitter, measuring the additional activity generated by Twitter's promotion. At the individual level, \citet{popularity-shocks-user} found that users who suddenly became popular increased their posting frequency and adapted their content to resemble what had initially driven their popularity. At the community level, \citet{increased-attention-github} analyzed how GitHub repositories changed after appearing on the platform's Trending Projects page, documenting substantial increases in attention alongside challenges in onboarding the influx of newcomers.
A related but less technical case comes from Reddit's r/NoSleep, studied by \citet{eternal-september}, which became a ``default'' subreddit automatically subscribed to by all new users. The authors interviewed the moderators of r/NoSleep and found that the community successfully adapted to a significant newcomer influx through strong moderator coordination, distributed moderation among community members, and the use of automated moderation tools.

All in all, our work extends \citet{chan-popular-audit}'s examination of \popular{} by shifting from algorithmic auditing using observational data alone to directly measuring user perceptions through an experimental approach. We focus on trending feeds because they are key cultural focal points---places where users encounter content not because it is recent, but because it has already attracted substantial engagement from others. In doing so, trending feeds not only reflect culture but actively shape it, becoming cultural objects in their own right~\cite{trending-is-trending}.

\section{Experimental Design \& Operationalization} \label{sec:design}

In this section, we provide an overview of the experiment, including the tasks participants complete, and how we operationalize each research question and hypothesis.

\begin{figure*}[t]
    \centering
    \begin{subfigure}{\textwidth}
        \centering
        \includegraphics[width=0.6\textwidth]{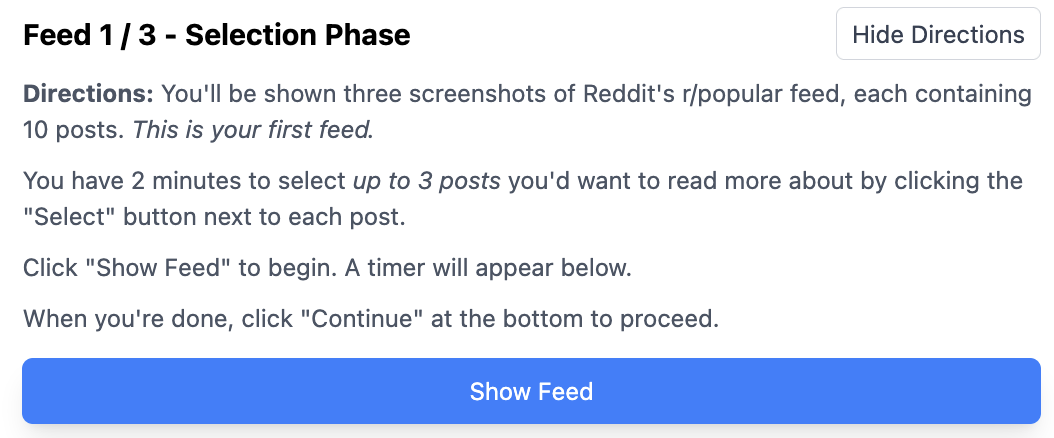}
        \caption{Instructions shown during the ``Selection Phase.''}
        \label{fig:top}
    \end{subfigure}
    
    \vspace{1em}
    
    \begin{subfigure}[c]{0.47\textwidth}
        \includegraphics[width=\textwidth]{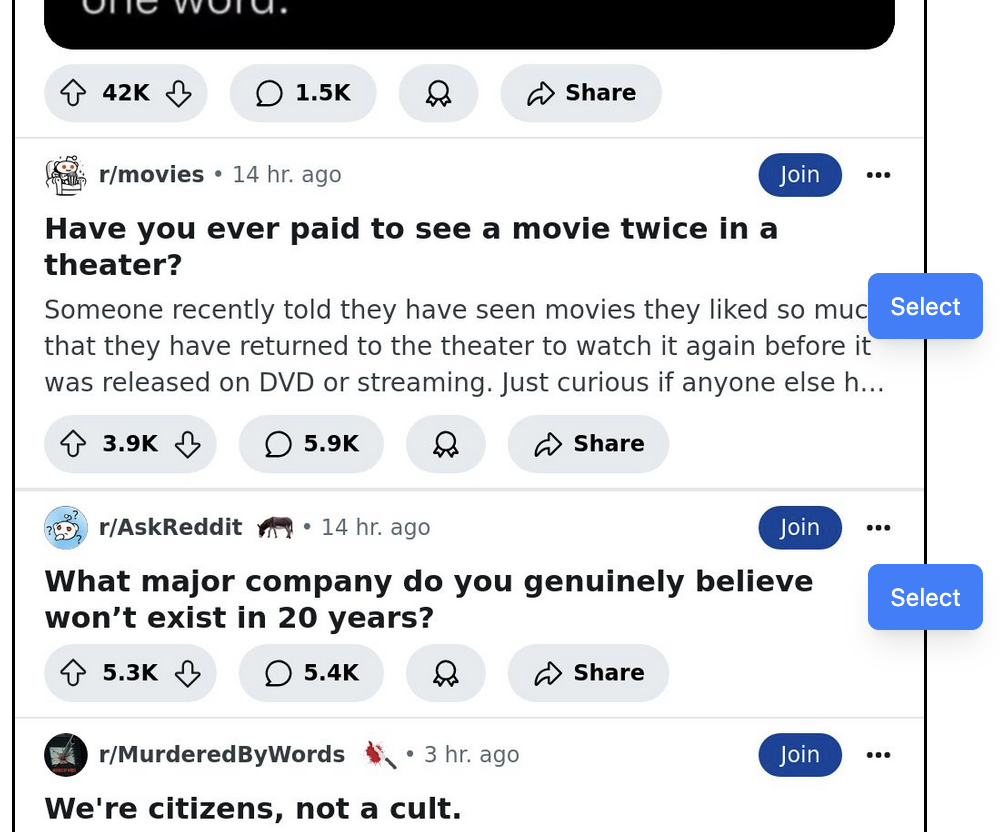}
        \caption{``Select'' button displayed with each post during ``Selection Phase.'' During the ``Rating Phase'' they turn into ``Rate'' buttons.}
        \label{fig:buttons}
    \end{subfigure}
    \hfill
    \begin{subfigure}[c]{0.47\textwidth}
        \centering
        \includegraphics[width=0.9\textwidth]{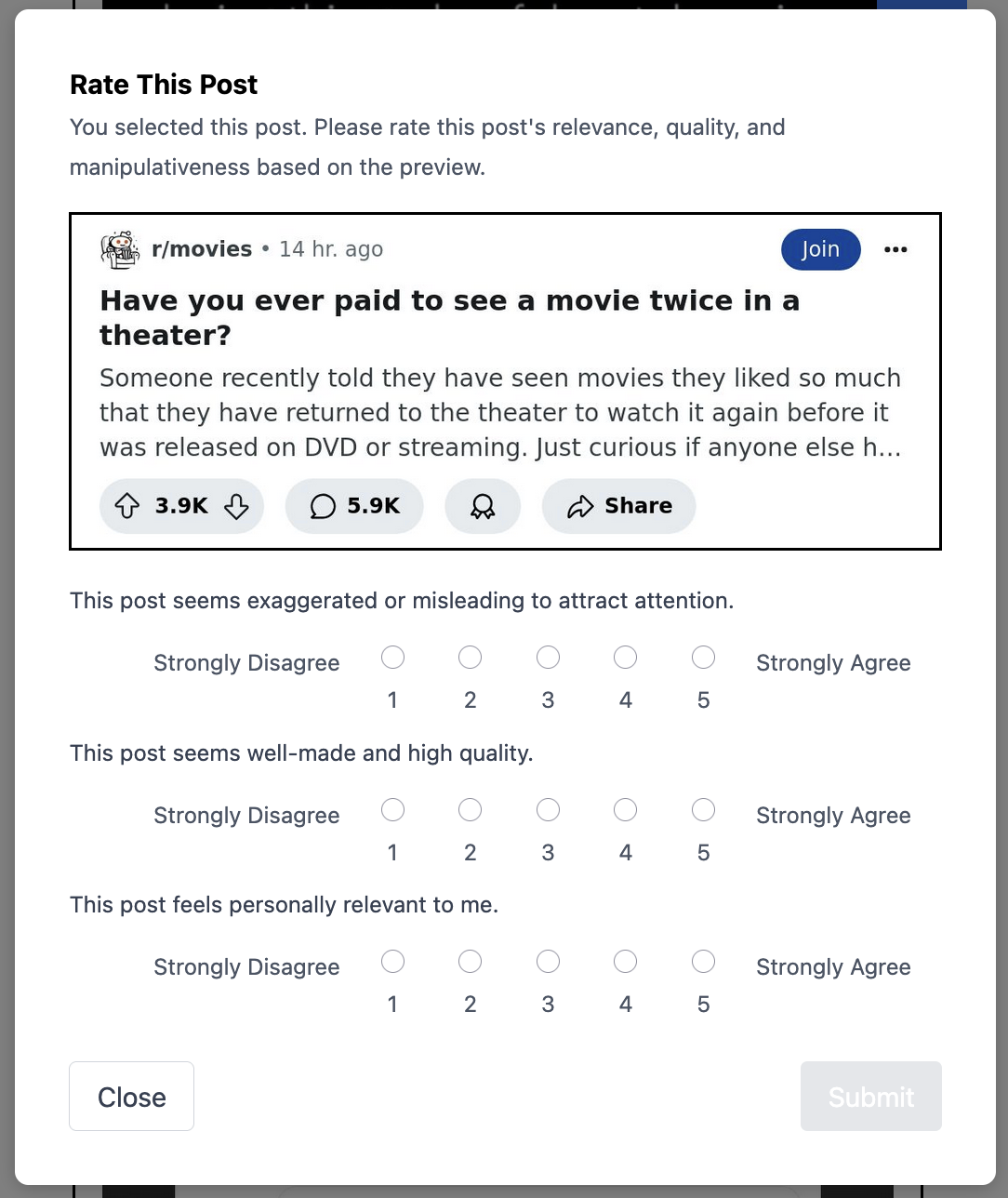}
        \caption{Pop-up window for rating posts by relevance, trustworthiness, and content quality.}
        \label{fig:popup}
    \end{subfigure}
    
    \caption{Interface elements from our web application that administers the experiment described in Section~\ref{sec:design}. To see how we rotate the same ten posts, see Appendix~\ref{fig:full-interface}.}
    \label{fig:interface}
    \Description{
        Three screenshots of the web application used in the experiment. The first screenshot displays the instructions for the ``Selection'' phase, where participants choose posts from r/popular's top 10. The second screenshot shows the top 10 posts, each with an adjacent ``Select'' button for marking chosen posts. The third screenshot shows the pop-up window where participants rate the selected posts.
    }
\end{figure*}

\subsection{Experiment Overview}

Our study involves three stages: (1) a pre-experiment questionnaire, (2) the feed experiment, and (3) an exit questionnaire.

Since this study focuses on Reddit, participants must have prior experience using Reddit. In addition to prior Reddit usage, we also require participants to be at least 18 years old and reside in the United States of America.

\subsubsection{Pre-Experiment Questionnaire}

After reading the consent form and providing their consent, participants complete a pre-experiment questionnaire. This includes two questions about their interests, which are used later in the analysis, and an attention check question (one of three administered across the experiment stages).

\begin{enumerate}
    \item Which of these topics are you generally interested in? Please select all that apply. You must choose at least one.
\end{enumerate}

The list of topics is drawn from Reddit's sidebar under the ``Topics'' tab and includes categories such as anime, business, games, and music. In total, there were 24 available topics.

Participants are then asked to list subreddits they visit frequently:

\begin{enumerate}
    \setcounter{enumi}{1}
    \item List the subreddits you visit most often. Try to list at least five.
\end{enumerate}

After completing these questions, participants proceed to the main portion of the experiment.

\subsubsection{Feed Experiment}

The feed experiment has two phases---\textit{selection} and \textit{rating}. In the \textit{selection phase}, participants view a screenshot of Reddit's \popular{} feed containing 10 posts. They have 2 minutes to select between one and three posts based on whether they would like to read more about them. The directions are shown in Figure~\ref{fig:top}.

Although participants view only a static screenshot, prior research~\cite{reddit-usage-data} indicates that 73\% of posts are voted on without reading the content. This suggests that static screenshots are sufficient for participants to determine whether they would engage with a post.

In the \textit{rating phase}, participants see the same feed with six posts---comprising their selections plus randomly chosen unselected posts---and rate each on perceived relevance, trustworthiness, and content quality using the following statements on a five-point Likert scale from ``strongly disagree'' to ``strongly agree'':

\begin{itemize}
    \item \textbf{Relevance:} This post feels personally relevant to me.
    \item \textbf{Trustworthiness/Manipulation:} This post seems exaggerated or misleading to attract attention. (reverse-coded)
    \item \textbf{Content Quality:} This post seems well-made and high quality.
\end{itemize}

The order of the statements is randomized for each participant to mitigate ordering effects. See Figure~\ref{fig:popup} for the interface element we used to record the ratings. Participants repeat these two phases for two additional screenshots---each containing a different set of 10 posts---before moving to the exit questionnaire.

These three perceptual dimensions (i.e., relevance, trustworthiness, and quality) were selected based on a literature review. \citet{trending-is-trending} find that participants often browse trending feeds to discover content they deem \textit{relevant}, frequently using engagement as a proxy. This raises a critical question: to what extent do trending feeds themselves shape perceptions of relevance? For example, could a platform's decision to highlight and rank a news story substantially influence how relevant it feels to viewers? The only prior work on relevance in social media, by \citet{munmun-relevance}, examined whether an alternative Twitter curation algorithm surfaced more relevant content than the platform's existing system. However, their study evaluated relevance in aggregate, whereas we investigate post-level relevance and how both ranking and engagement metrics shape it.

Our focus on perceived \textit{trustworthiness} reflects the frequency of ``breaking news'' posts on \popular{}. Such posts are consequential because they can facilitate the spread of misinformation, a central concern in social media research~\cite{tweeting-believing, synthesized-social-signals}. Prior work on search engines shows that users often place undue trust in ranked results~\cite{ranking-versus-reputation, biased-search-debated-topics}, raising the possibility that trending feeds exert similar influence. Because \popular{} features news alongside other topical content, its rankings risk serving as platform endorsement signals. We therefore include \textit{trustworthiness} as a key perceptual dimension.

Unlike relevance or trustworthiness, \textit{content quality} is a broad construct encompassing many factors~\cite{content-quality-frameworks} from engagement levels to title phrasing. Since trending feeds are intended to showcase the platform's ``best'' content, we ask whether this intent is reflected in users' perceptions, and whether rank and engagement bias those judgments. If so, algorithms could exacerbate engagement inequality, with highly ranked posts attracting disproportionate attention and reinforcing a rich-get-richer dynamic.

\subsubsection{Exit Questionnaire}

The first three exit questions ask participants to reflect on their selection choices. The first question directly addresses RQ1:

\begin{enumerate}
    \item How did you determine which posts you selected? Select all that apply.
\end{enumerate}

Participants could choose from the following options: (1) position of the post on the feed, (2) subreddit of the post, (3) number of comments, (4) number of upvotes, (5) content of the post, plus an open-text ``Other'' option. The options were presented in random order---with the exception of ``Other,'' which was always listed last---to prevent ordering effects.

The next two questions provide random examples of one post they \textit{did} select and one they \textit{did not} select:

\begin{enumerate}
    \setcounter{enumi}{1}
    \item You selected the above post. Which factors influenced your choice? Select all that apply.
    \item You did not select the above post. Which factors influenced that decision? Select all that apply.
\end{enumerate}

For the second question, options include: (1) ``This post felt personally relevant to me,'' (2) ``This post did not seem exaggerated or misleading,'' (3) ``This post seemed well-made and high quality,'' plus an open-text ``Other'' option. The third question uses the inverse wording for each statement. Both of the above questions have their option order randomized.

Participants also rate how realistic the screenshots of \popular{} were:

\begin{enumerate}
    \setcounter{enumi}{3}
    \item How likely are you to encounter content similar to what you saw during the experiment? (Five-point Likert scale from ``highly unlikely'' to ``highly likely'')
\end{enumerate}

Finally, participants provide demographic information, including gender, age, and highest educational attainment.

\subsection{Experimental Stimuli: Creating Controlled Variations in Screenshots of \popular{}}

Participants interact primarily with screenshots of \popular{}, which serve as the experimental stimuli. We create controlled variations of a given screenshot in two ways:

First, posts are rotated in a circular manner rather than shuffled. This reduces the number of conditions from 10! (if shuffled) to 10 while preserving neighboring posts, minimizing potential neighbor effects. For example, a post flanked by high-quality neighbors may be perceived differently; rotation keeps these neighbor effects consistent. To see an example of how posts are rotated, see Figure~\ref{fig:full-interface} in Appendix~\ref{sec:full-interface}.

Second, some screenshots display the number of comments and upvotes, while others hide them. This allows us to assess whether these forms of social proof, as defined by \citet{social-proof-definition}, influence perceptions of relevance, trustworthiness, and content quality. Each participant sees either all screenshots with social proof or all without to avoid within-subject contamination. To see how a post looks with and without social proof, see Figure~\ref{fig:social-proof} in Appendix~\ref{sec:social-proof}.

Thus, the experiment follows a $10 \times 2$ factorial design, with three repeated measures per participant corresponding to the three screenshot variations they encounter.

To construct the stimuli, we developed a JavaScript program that retrieves Reddit's \popular{} feed and systematically modifies the DOM to reorder posts and toggle social proof metrics. We define a \textit{snapshot} as the set of 10 posts appearing on \popular{} at a given time. For each snapshot, the program generates multiple \textit{screenshots}, each representing a different experimental configuration of those same 10 posts. In our case, recording one snapshot required 20 screenshots to capture all desired variations.  

Snapshots were collected every two hours from July 15 to 29, 2025, yielding 155 in total. From these, we randomly sampled five snapshots, and each participant was assigned three. Because each snapshot consists of 10 posts, this produced 50 unique posts drawn from 36 subreddits. These included four content types: images \n{29}, videos \n{11}, links \n{6}, and text \n{4}. The five sampled snapshots were captured on July 15, 2025 (4:00 PM), July 17 (8:00 AM), July 19 (12:00 PM), July 21 (12:00 PM), and July 25 (8:00 PM).

\subsection{Operationalizing Research Questions \& Hypotheses}

Having outlined the experiment, we now describe how the study design addresses each research question and hypothesis.

\subsubsection{RQ1: Post Factors Influencing Selection}

To examine which factors influence post selection on \popular{}, we combine self-report measures with behavioral modeling. First, participants indicate in the exit questionnaire which factors guided their choices. These responses will be analyzed quantitatively using Cochran's Q test, with McNemar's test for post hoc pairwise comparisons, and qualitatively through open-text responses to the ``Other'' option. Second, we estimate the likelihood of a post being selected within a given feed and rank using a generalized linear mixed model (GLMM). The full specification of this model, including the regression equation, is provided in Section~\ref{sec:data-analysis}. Together, these approaches allow us to assess both perceived and observed drivers of post selection.

\subsubsection{H2: Influence of Relevance, Trustworthiness, \& Content Quality on Selection \& Non-Selection}

To evaluate whether perceived relevance, trustworthiness, and content quality are primary factors shaping selection decisions, we use participants' responses to exit questionnaire questions (2) and (3). In these items, participants are shown an example of a post they selected and one they did not select, and asked to indicate their reasoning by choosing from predefined options corresponding to relevance, trustworthiness, and content quality. An ``Other'' option is also available; these responses will be analyzed qualitatively to identify additional themes when the three main perceptions are not cited. To test for differences in the frequency of selections across the three predefined options, we will use Cochran's Q test, with McNemar's test for post hoc pairwise comparisons.

\subsubsection{RQ3: Demographic Influences on Selection \& Perceptions}

We collect participants' age, gender, and highest educational attainment through multiple-choice questions. Each of these variables is included in the GLMMs as random intercepts. 

% In addition to the GLMM modeling selection likelihood
In addition to the GLMM that models likelihood of selection, we train three additional GLMMs to estimate perceived relevance, trustworthiness, and content quality based on ratings provided for selected and non-selected posts during the experiment. We refer to these as the \textit{rating} models. By analyzing the random intercepts associated with demographic variables, we can determine whether these factors significantly influence selection behaviors and perceptions of content.

\subsubsection{RQ4: Influence of Content Preferences on Selection \& Perceptions}

Participants' content preferences are collected in the pre-experiment questionnaire by asking them to select topics of interest and provide subreddits they browse. Both the selected topics and subreddits are included as predictors in the selection and rating GLMMs. By examining the model coefficients, we can quantify how content preferences shift engagement and perceptions of posts on \popular{}. Further details on how these responses are incorporated into the models are provided in Section~\ref{sec:data-analysis}.

\subsubsection{H5: Influence of Rank on Selection \& Perceptions}

All GLMMs include post rank as a feature. This allows us to test whether rank significantly influences selection or perceptions of content. Additionally, the exit questionnaire includes ``position of the post'' as a possible reason for selection. Comparing participants' self-reported awareness with model estimates of rank effects enables us to assess whether rank exerts an implicit influence on selection and perceptions.

\subsubsection{H6: Influence of Social Proof on Selection \& Perceptions}

We manipulate the presence of social proof (i.e., number of comments and upvotes) for posts shown to participants. Both the selection and rating GLMMs include a predictor for whether social proof is shown, as well as the actual number of comments and upvotes to capture its degree. This allows us to assess not only whether the presence of social proof influences selection and perceptions, but also how the \textit{degree} of social proof affects these outcomes.

% BLUE TEXT INDICATE THESE ARE VARIABLES.
\newcommand{\compensation}{\textcolor{black}{\$2.50 USD}}
\newcommand{\crowdworkplatform}{\textcolor{black}{Prolific}}
\newcommand{\numcompleted}{\textcolor{black}{585}}
\newcommand{\numfailedattention}{\textcolor{black}{7}}
\newcommand{\numfailedtech}{\textcolor{black}{15}}

% Man                  322
% Woman                250
% Non-binary            10
% Prefer not to say      3
\newcommand{\percentagemale}{\textcolor{black}{55.04\% ($n=322$)}}
\newcommand{\percentagefemale}{\textcolor{black}{42.74\% ($n=250$)}}
\newcommand{\percentagenonbinary}{\textcolor{black}{1.71\% ($n=10$)}}
\newcommand{\percentagenoanswer}{\textcolor{black}{0.51\% ($n=3$)}}
\newcommand{\mediancompletiontime}{\textcolor{black}{13 minutes and 9 seconds ($\mu=14m\ 33s$, $\sigma=7m\ 11s$)}}
% https://osf.io/2qfp5

\section{Methods} \label{sec:methods}

This study was granted an IRB exemption by our institution on July 10, 2025, and preregistered on OSF.\footnote{\url{https://osf.io/2qfp5}}
% \footnote{Censored for review: \url{https://osf.io/XXXXX}}

\subsection{Experiment Deployment}

We piloted the study with departmental colleagues and a paid sample of 10 participants on Prolific, who were compensated \compensation{} for their time. These pilots informed our estimated study duration, improved the clarity of instructions and survey wording, and guided refinements to the user interface of the custom tool used to administer the experiment.

The experiment was deployed through a custom web application built with React, FastAPI, and MongoDB. Participants were recruited via Prolific, which passed custom URL parameters that uniquely identified them for compensation purposes. Using Prolific's screening tools, we restricted eligibility to participants who (1) were at least 18 years old, (2) resided in the United States, and (3) reported prior Reddit use. The study had an estimated completion time of 12 minutes, and participants were compensated \compensation{}.

We recruited \numparticipants{} participants between August 9 and August 22, 2025, until funds were depleted. Of these, \numcompleted{} participants completed the study without failing attention checks; \numfailedattention{} failed two or more of the three embedded attention checks and were excluded (and not compensated, per Prolific's policy for surveys exceeding 10 minutes);\footnote{Prolific requires participants to fail at least two attention checks before compensation may be withheld for surveys longer than 10 minutes.} and \numfailedtech{} did not complete the survey due to technical issues. These latter participants were compensated, as the source of error could not be determined.

Among the completed sample, \percentagemale{} identified as men, \percentagefemale{} as women, \percentagenonbinary{} as non-binary, and \percentagenoanswer{} declined to answer. The median completion time was \mediancompletiontime{}. Additional demographic details are reported in Appendix~\ref{sec:demographics}.

\subsection{Data Analysis} \label{sec:data-analysis}

In addition to the Cochran's Q tests and McNemar's test for post hoc analysis we will use for the multiple-choice questions, we train two types of GLMMs to answer/validate our research questions and hypotheses. In this section, we provide details about these models and how we represent factors like rank, social proof, and participant content preferences.

\subsubsection{Selection Model} \label{model:selection}

To estimate the likelihood of a post being selected, we use a Bayesian mixed-effects logistic regression implemented in \texttt{bambi} (BAyesian Model-Building Interface in Python)\footnote{\url{https://bambinos.github.io/bambi/}} which builds on top of \texttt{PyMC}\footnote{\url{https://www.pymc.io/welcome.html}} to fit hierarchical models. The regression equation defining the selection model is shown below:

\iffalse
\begin{equation*}
\begin{aligned}
\text{logit}\big(\Pr(\text{selected}_{i, p} = 1)\big) &= 
    \underbrace{
        \beta_0 + \beta_1\ \text{shown\_proof}(p) + \beta_2\ \text{rank}(i) + \text{shown\_proof}(p) \cdot \beta_3[\text{rank}(i)]
    }_{\text{main effects \& rank interaction}} \\
    %
    &\quad + \underbrace{
    \beta_4\,(\text{shown\_proof}(p)\cdot \text{comments}(i))
        + \beta_5\,(\text{shown\_proof}(p)\cdot \text{upvotes}(i))
    }_{\text{degree of social proof}} \\
    &\quad + \underbrace{
    \beta_6\,\text{is\_subreddit\_of\_interest}(i, p)
        + \beta_7\,\text{is\_topic\_of\_interest}(i, p)
    }_{\text{content preferences effects}} \\
    %
    &\quad + \underbrace{
        u_{\text{age}}(p) + u_{\text{gender}}(p) + u_{\text{education}}(p)
        + u_{\text{post}}(i)
    }_{\text{demographic \& post random effects}} \\
    %
    % &\quad + \underbrace{
    %     u_{\text{participant}}(p) + u_{\text{snapshot}}(s) + u_{\text{post}}(i)
    %     + u_{\text{subreddit}}(i) + u_{\text{content\_type}}(i)
    % }_{\text{other random effects}}
\end{aligned}
\end{equation*}
\fi

\begin{equation*}
\begin{aligned}
\text{logit}\big(\Pr(\text{selected}_{i, p} = 1)\big) &= 
    \beta_0 + \beta_1\,\text{shown\_proof}(p) \\
    &\quad + \beta_2\,[\,\text{rank}(i)\,] \\
    &\quad + \beta_3\,[\,\text{rank}(i)\,]\cdot\text{shown\_proof}(p) \\
    &\quad + \beta_4\,\text{shown\_proof}(p)\cdot \text{comments}(i) \\
    &\quad + \beta_5\,\text{shown\_proof}(p)\cdot \text{upvotes}(i) \\
    &\quad + \beta_6\,\text{is\_subreddit\_of\_interest}(i, p) \\
    &\quad + \beta_7\,\text{is\_topic\_of\_interest}(i, p) \\
    &\quad + u_{\text{age}}(p) + u_{\text{gender}}(p) \\
    &\quad + u_{\text{education}}(p) + u_{\text{post}}(i)
\end{aligned}
\end{equation*}

% In this model, post $i$ and participant $p$ are the primary inputs. The snapshot $s$ identifies which of the five snapshots (i.e., ten-post corpora) a participant could have viewed. The variable \texttt{shown\_proof} is coded as 0 (no proof) or 1 (proof) based on the participant's assigned condition.

In this model, the primary inputs are post $i$ and participant $p$. The variable \texttt{shown\_proof} is coded as 0 (no social proof) or 1 (social proof), depending on the participant's condition. 

In addition to the main effects for social proof ($\beta_1$) and rank ($\beta_2$), we also include an interaction term ($\beta_3$) between rank and \texttt{shown\_proof}, allowing the model to capture whether the effect of rank on post selection differs depending on the presence of social proof. This enables us to determine if posts at specific ranks are more or less likely to be selected when social proof is shown.

The terms $\beta_4$ and $\beta_5$ capture how the degree of social proof (i.e., post-level engagement metrics) influences selection odds. The terms $\beta_6$ and $\beta_7$ capture user-level interests in the subreddit and topic, respectively. Lastly, the $\mu$ terms are intercept terms for our demographic variables and one term specifically for the post itself.

\newcommand{\interestedsubreddit}{\texttt{is\_sub\-reddit\_of\_interest}}
\newcommand{\interestedtopic}{\texttt{is\_topic\_of\allowbreak\_interest}}

Finally, to account for content preferences, we include \interestedsubreddit{} and \interestedtopic{}. The variable \interestedsubreddit{} indicates whether the participant reported browsing the subreddit in the pre-experiment questionnaire, while \interestedtopic{} captures whether the participant selected the topic associated with the post. Topics were labeled by two authors independently for all 50 posts across the five snapshots, with disagreements resolved through discussion.

\subsubsection{Rating Models} \label{model:rating}

To estimate perceived relevance, trustworthiness, and content quality, we use Bayesian mixed-effects ordinal regression models implemented in \texttt{bambi}. These models operate in a latent space ($z$) and learn cut points that map the latent variable $z$ to the observed ratings on the Likert scale.
The model uses the same inputs as the selection model---post $i$ and participant $p$.
The complete formula for each perception is:

\iffalse
\begin{equation*}
\begin{aligned}
\text{logit}\Big(\Pr(\text{perception}_{i,p} \le k)\Big) &= 
    \underbrace{
    \beta_0 + \beta_1 \text{shown\_proof}(p) + \beta_2 \text{rank}(i)
        + \text{shown\_proof}(p) \cdot \beta_3[\text{rank}(i)]
    }_{\text{main effects \& rank interaction}} \\
    %
    &\quad + \underbrace{
        \beta_4\,(\text{shown\_proof}(p)\cdot \text{comments}(i))
        + \beta_5\,(\text{shown\_proof}(p)\cdot \text{upvotes}(i))
    }_{\text{degree of social proof}} \\
    %
    % &\quad + \underbrace{
    %     \beta_6\,\text{selected}(i,p)
    % }_{\text{effect of selection}} \\
    %
    &\quad + \underbrace{
        \beta_6\,\text{is\_subreddit\_of\_interest}(i, p) + \beta_7\,\text{is\_topic\_of\_interest}(i, p)
    }_{\text{content preferences effects}} \\
    %
    &\quad + \underbrace{
        u_{\text{age}}(p) + u_{\text{gender}}(p) + u_{\text{education}}(p)
        + u_{\text{post}}(i)
    }_{\text{demographic \& post random effects}} \\
    %
    % &\quad + \underbrace{
    %     u_{\text{participant}}(p) + u_{\text{snapshot}}(s) + u_{\text{post}}(i) + u_{\text{subreddit}}(i) + u_{\text{content\_type}}(i)
    % }_{\text{other random effects}}
\end{aligned}
\end{equation*}
\fi

\begin{equation*}
\begin{aligned}
\text{logit}\Big(\Pr(\text{percept}_{i,p} \leq k)\Big) &= 
    \beta_0 + \beta_1\,\text{shown\_proof}(p) \\
    &\quad + \beta_2\,[\,\text{rank}(i)\,] \\
    &\quad + \beta_3\,[\,\text{rank}(i)\,]\cdot\text{shown\_proof}(p) \\
    &\quad + \beta_4\,\text{shown\_proof}(p)\cdot \text{comments}(i) \\
    &\quad + \beta_5\,\text{shown\_proof}(p)\cdot \text{upvotes}(i) \\
    &\quad + \beta_6\,\text{is\_subreddit\_of\_interest}(i, p) \\
    &\quad + \beta_7\,\text{is\_topic\_of\_interest}(i, p) \\
    &\quad + u_{\text{age}}(p) + u_{\text{gender}}(p) \\
    &\quad + u_{\text{education}}(p) + u_{\text{post}}(i)
\end{aligned}
\end{equation*}

This model allows us to estimate how the presence and degree of social proof, post rank, and demographic variables influence participants' perceptions---abbreviated ``percept'' in the equation above---of posts across the three dimensions, while accounting for hierarchical structure in the data. The cut points learned by the model map the latent variable to the ordinal Likert ratings for each perception.

\newcommand{\rsquaredselection}{\textcolor{black}{$0.346$}}
\newcommand{\rsquaredrelevance}{\textcolor{black}{$0.365$}}
\newcommand{\rsquaredtrust}{\textcolor{black}{$0.361$}}
\newcommand{\rsquaredcontentquality}{\textcolor{black}{$0.359$}}

% --- Findings Answers ---
\newcommand{\rqoneanswer}{Post content and subreddit were the strongest drivers of selection on \popular{}.}
\newcommand{\htwoanswer}{Relevance most strongly guided selection and non-selection. Trustworthiness mattered for non-selections. Outside these, being interesting, funny, or curiosity-sparking also influenced choices.}
\newcommand{\rqthreeanswer}{Younger participants (18--24) saw posts as more relevant; older participants viewed them as more manipulative. Ages 35--64 rated posts as lower quality than 25--34 year olds.}
\newcommand{\rqfouranswer}{Interest in the topic or subreddit sharply boosted selection odds, raising perceived relevance and quality while lowering perceived manipulation.}
\newcommand{\hfiveanswer}{Lower-ranked posts were less likely to be selected than rank~1, but rank did not affect perceptions.}
\newcommand{\hsixanswer}{Neither social proof nor its degree significantly influenced selection or perceptions.}

% --- Findings Table ---
\begin{table*}[t]
    \renewcommand{\arraystretch}{1.3}
    \centering
    \small
    \caption{Summary of findings for each research question and hypothesis.}
    \label{tab:findings-summary}
    \begin{tabular}{p{4.5cm} p{9cm}}
        \toprule
        \textbf{Research Question / Hypothesis} & \textbf{Findings} \\
        \midrule
        \textbf{RQ1:} Post Factors Influencing Selection & \rqoneanswer \\
        \textbf{H2:} Influence of Relevance, Trustworthiness, \& Content Quality & \htwoanswer \\
        \textbf{RQ3:} Demographic Influences on Selection \& Perceptions & \rqthreeanswer \\
        \textbf{RQ4:} Influence of Content Preferences on Selection \& Perceptions & \rqfouranswer \\
        \textbf{H5:} Influence of Rank on Selection \& Perceptions & \hfiveanswer \\
        \textbf{H6:} Influence of Social Proof on Selection \& Perceptions & \hsixanswer \\
        \bottomrule
    \end{tabular}
\end{table*}

\section{Findings}

Before addressing each research question and hypothesis, we first summarize the performance of the two sets of GLMMs used in this study. The selection model described in Section~\ref{model:selection} yielded an $R^2$ of \rsquaredselection{}, indicating that a substantial portion of the variance in selection behavior remains unexplained. For the rating models described in Section~\ref{model:rating}, the $R^2$ values were \rsquaredrelevance{}, \rsquaredtrust{}, and \rsquaredcontentquality{} for relevance, trustworthiness/manipulation, and content quality, respectively.

Table~\ref{tab:selection-frequencies} reports the number of times participants selected a post in each social proof condition as well as overall. Table~\ref{tab:selection-reasons} summarizes participants’ self-reported reasons for their selections on \popular{}, collected during the exit questionnaire. Likewise, Table~\ref{tab:perception-reasons} presents participants’ reasons for either selecting or not selecting an example post.

Finally, Table~\ref{tab:findings-summary} provides a concise overview of the findings associated with each research question and hypothesis.

\begin{table*}[t]
    \centering
    \begin{minipage}{0.49\textwidth}
        \centering
        \caption{Frequencies of reasons participants reported as influencing their decision to select posts on \popular{}. Percentages are calculated relative to the number of participants ($n=585$).}
        \begin{tabular}{lrr}
            \toprule
            \textbf{Reason} & $n$ & \% \\
            \midrule
            Content   & 550 & 94.0\% \\
            Subreddit & 178 & 30.4\% \\
            Comments  & 35  & 6.0\% \\
            Upvotes   & 31  & 5.3\% \\
            Other     & 30  & 5.1\% \\
            Position  & 27  & 4.6\% \\
            \bottomrule
        \end{tabular}
        \label{tab:selection-reasons}
    \end{minipage}
    \hfill
    \begin{minipage}{0.49\textwidth}
        \centering
        \caption{Frequencies of perceptions participants reported as influencing their decision to either select or not select posts on \popular{}. Percentages are calculated relative to the number of participants ($n=585$).}
        \begin{tabular}{lrrrr}
            \toprule
            \textbf{Reason} 
            & \multicolumn{2}{c}{\textbf{Selection}} 
            & \multicolumn{2}{c}{\textbf{Non-Selection}} \\
            \cmidrule(lr){2-3} \cmidrule(lr){4-5} 
            & $n$ & \% & $n$ & \% \\
            \midrule
            Relevance        & 268 & 45.8\% & 373 & 63.8\% \\
            Trustworthiness  & 153 & 26.2\% & 167 & 28.5\% \\
            Content Quality  & 149 & 25.5\% & 113 & 19.3\% \\
            Other            & 191 & 32.6\% & 84  & 14.4\% \\
            \bottomrule
        \end{tabular}
        \label{tab:perception-reasons}
    \end{minipage}
\end{table*}

\begin{table*}[ht]
    \centering
    
    % --- Selection Reasons ---
    \begin{minipage}{0.9\textwidth}
        \centering
        \caption{Pairwise McNemar's tests for factors influencing participants' post selection decisions on \popular{}. 
        Content was the strongest self-reported factor, followed by subreddit, though content clearly dominated.}
        \label{tab:mcnemar-post-factors}
        \begin{tabular}{lcccccc}
            \toprule
            & \textbf{Position} & \textbf{Content} & \textbf{Upvotes} & \textbf{Comments} & \textbf{Subreddit} & \textbf{Other} \\
            \midrule
            \textbf{Position}   & --   &      &      &      &      &     \\
            \textbf{Content}    & >    & --   &      &      &      &     \\
            \textbf{Upvotes}    & ns   & <    & --   &      &      &     \\
            \textbf{Comments}   & ns   & <    & ns   & --   &      &     \\
            \textbf{Subreddit}  & >    & <    & >    & >    & --   &     \\
            \textbf{Other}      & ns   & <    & ns   & ns   & <    & --  \\
            \bottomrule
        \end{tabular}
    \end{minipage}
    
    \vspace{.75em}

    \caption*{\footnotesize 
        Entries indicate results of pairwise McNemar's tests. 
        ns = not significant; 
        $>$ indicates a significant relationship ($p<0.05$, Bonferroni-adjusted) favoring the \emph{row} factor; 
        $<$ indicates a significant relationship favoring the \emph{column} factor.
    }

    \vspace{.75em}

    % --- Perceptions ---
    \begin{minipage}{0.9\textwidth}
        \centering
        \caption{Pairwise McNemar's tests for perceptions influencing participants' selection and non-selection of posts on \popular{} (selection / non-selection). 
        Relevance dominated both selection and non-selection, with trustworthiness second for non-selection.}
        \label{tab:mcnemar-perceptions}
        \label{tab:}
        \begin{tabular}{lcccc}
            \toprule
            & \textbf{Relevance} & \textbf{Trustworthiness} & \textbf{Content Quality} & \textbf{Other} \\
            \midrule
            \textbf{Relevance}       & --        &       &       &      \\
            \textbf{Trustworthiness} & < / <     & --    &       &      \\
            \textbf{Content Quality} & < / <     & ns / < & --   &      \\
            \textbf{Other}           & < / <     & ns / < & ns / ns & -- \\
            \bottomrule
        \end{tabular}
    \end{minipage}

    \vspace{.75em}

    \caption*{\footnotesize 
        Entries indicate results of pairwise McNemar's tests. 
        ns = not significant; 
        $>$ indicates a significant relationship ($p<0.05$, Bonferroni-adjusted) favoring the \emph{row} factor; 
        $<$ indicates a significant relationship favoring the \emph{column} factor.
    }
\end{table*}

\subsection{RQ1: Post Factors Influencing Selection}

% personal_interest/curiosity/caring - 15
% funny/entertainment - 3
% content - 3
% comments - 3
% lack_of_options - 2
% post_title - 4
% read_more - 3
% visual_media - 3
% credibility - 2
% --- Total "Other" responses: 30 ---

The exit questionnaire asked participants: ``How did you determine which posts you selected?'' Table~\ref{tab:selection-reasons} reports the frequency of responses. The most frequently cited factor was the content of the post~(65.34\%, $n=550$), followed by the subreddit in which the post appeared~(21.15\%, $n=124$).

To test for significant differences between these response frequencies, we first conducted a Cochran's $Q$ test, a non-parametric omnibus test of differences across related proportions. The test was significant at the Bonferroni-adjusted threshold of $p < 0.05$, so we proceeded with pairwise McNemar's tests to identify specific contrasts. The results of these pairwise comparisons are presented in Table~\ref{tab:mcnemar-post-factors}.

Table~\ref{tab:mcnemar-post-factors} shows that a post's content dominated all other reasons for selection, followed by the subreddit of the post. In contrast, differences between our manipulated variables, rank and the two social proof cues (number of upvotes and number of comments), were not statistically significant.

\subsubsection{Qualitative Themes from ``Other'' Responses}

Finally, 30 participants provided write-in responses under the ``Other'' option. We manually coded these to identify additional themes. The most common theme was personal interest in the topic of the post~(50.00\%, $n=15$), e.g., ``topics I was personally interested in.'' Other less frequent responses included the post title~($n=4$), posts being funny~($n=3$), interest in the comment section~($n=3$, e.g., ``The posts that I deemed most likely to have interesting comments.''), content itself despite an existing content option~($n=3$), the image~($n=3$), and the source of a linked news article~($n=2$). Overall, the main factor not captured by the closed-ended options was participants' personal interest in the topic, which---unlike the post's subreddit, content, or engagement metrics---was not a component of the post itself and therefore was not included as an option.

\subsection{H2: Influence of Relevance, Trustworthiness, \& Content Quality}

In addition to asking participants broadly about the factors that influenced their post selections on \popular{}, we also presented each participant with a post they did and did not select and asked them to explain their decisions using the three perceptions: relevance, trustworthiness, and content quality. Participants could also select an open-ended ``Other'' option if none of these applied. The frequencies of each selected option are reported in Table~\ref{tab:perception-reasons}.

As with the post factors question, we first conducted a Cochran's $Q$ test, which was significant at the Bonferroni-adjusted threshold ($p<0.05$). We then conducted pairwise McNemar's tests, the results of which are shown in Table~\ref{tab:mcnemar-perceptions}. Overall, relevance (``This post [felt/did not feel] personally relevant to me'') emerged as the primary reason participants gave for both selection~(45.8\%, $n=268$) and non-selection decisions~(63.8\%, $n=373$)---the prominence of relevance compared to all other reasons was significant per our pairwise McNemar's tests in Table~\ref{tab:mcnemar-perceptions}. Trustworthiness (``This post [did not seem/seemed] exaggerated or misleading to attract attention'') was the second most common cited perceptions, however, there were more ``Other'' responses for the selection decisions~(32.6\%, $n=191$). Thus, trustworthiness did not significantly differ from either content quality or ``Other'' for selection decisions, but it was significantly more for non-selection when compared to content quality and ``Other.''

\subsubsection{Qualitative Themes from ``Other'' Responses}

Of note is the large number of ``Other'' responses~(32.6\%, $n=191$) for the selection question and for the non-selection question~(14.4\%, $n=84$). As with the post factors analysis, we manually coded these responses to identify themes.

% interesting - 46
% humor/funny/entertaining - 38
% comments - 32
% curiosity - 23
% content_specific - 16
% animals - 14
% read_more - 13
% lack_of_options - 9
% image/videos - 9
% other - 5
% personal_reason - 5
% wholesome/lighthearted - 4
% topical - 4
% exaggerated - 3
% title - 1
% relatable - 1
% verification - 1
% subreddit - 1
% --- Total "Other" responses: 191 ---

For the 191 ``Other'' responses describing why participants selected an example post, the most frequent themes were that the post was interesting~($n=46$), funny~($n=38$), or sparked curiosity~($n=23$). Participants also mentioned selecting posts to see the comments~($n=32$), or for more specific reasons such as unique content features~($n=16$, e.g., ``it sounded like a cute story''), the presence of animals~($n=14$, ``I was curious about the cat''), engaging media formats like videos or image galleries ($n=9$, ``i was interested in watching the video''), personal connections to the post~($n=5$), or topical relevance~($n=4$). In short, beyond relevance and trustworthiness, participants often selected posts because they were interesting, funny, or curiosity-inducing.

% political_fatigue - 24
% not_interesting_enough - 12
% unnecessary_click - 11
% already_read - 8
% sad - 5
% insults - 5
% untrustworthy - 5
% better_candidates - 4
% time_expired - 3
% need_more_info - 3
% other - 10 <---  I wasn't able to bin these. The responses were non-sense, e.g., ``The picture was worth 1000 words.''
% --- Total "Other" responses: 84 ---

For the 84 ``Other'' responses explaining participants non-selection decisions, the most common theme was political fatigue~($n=24$), with participants noting, for example, that they were ``currently avoiding most political content'' or that ``I basically see this kind of news on reddit every day, it gets tiring.'' Other reasons included that the post was not interesting enough to click~($n=12$), that participants felt they got everything they needed without clicking through~($n=11$), that the post was too sad~($n=5$), or needed more information to decide~($n=3$). A smaller number reported not selecting because they had already seen the post~($n=8$).

Overall, the ``Other'' responses suggest that participants' choices were also shaped by perceived ``interestingness,'' humor, and curiosity, factors not explicitly included in our predefined options. In addition, the prevalence of political fatigue highlights an important contextual influence during the period in which this study was conducted.

\subsection{RQ3: Demographic Influences on Selection \& Perceptions}

\begin{table}
\centering
\caption{
    Each rating model maps latent space $z$ onto the Likert scale using threshold cut points. The table reports the mean locations of these cut points for each perception and the distances between them. For example, $-0.642 < z < 0.292$ corresponds to a rating of 2 for relevance, a span of 0.934. These distances show how much latent change is needed to move between adjacent Likert categories.
}
\begin{tabular}{lrrrr}
    \toprule
     & \textbf{1/2} & \textbf{2/3} & \textbf{3/4} & \textbf{4/5} \\
    \midrule
    \textbf{Relevance}       & -0.642 &  0.292 &  1.285 &  2.630 \\
    \textit{Distance}        &        &  0.934 &  0.993 &  1.345 \\ \midrule
    \textbf{Manipulation}    & -1.046 &  0.008 &  0.967 &  2.134 \\
    \textit{Distance}        &        &  1.054 &  0.959 &  1.167 \\ \midrule
    \textbf{Content Quality} & -2.070 & -0.776 &  0.788 &  2.340 \\
    \textit{Distance}        &        &  1.294 &  1.564 &  1.552 \\
    \bottomrule
\end{tabular}
\label{tab:perception-cutpoints}
\end{table}

% TODO: Effect size?

Sections~\ref{model:selection} and \ref{model:rating} describe how we modeled age, gender, and educational attainment as categorical predictors, using the most common categories as reference groups: men, ages 25--34, and bachelor's degree holders. For distributions across categories, see Appendix~\ref{sec:demographics}.

\subsubsection{Selection}

Demographic variables did not significantly influence which posts participants selected. For example, women compared to men had a $3.80\%$ higher odds of selecting a post \onlyhdi{-3.29\%}{11.58\%}, but the HDI includes zero, indicating uncertainty in the effect direction. Age and education showed similarly null effects.

\begin{figure*}
    \centering
    \includegraphics[width=0.78\linewidth]{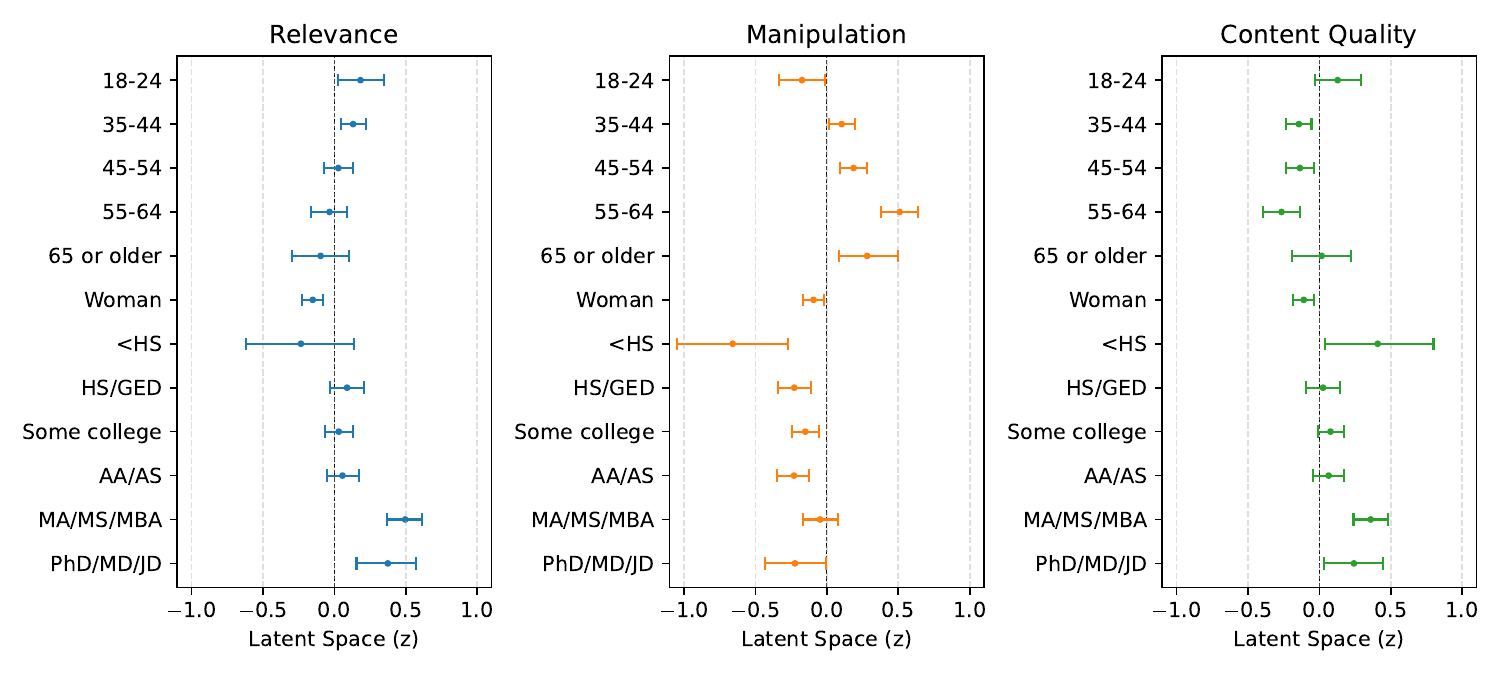}
    \caption{Effects of demographic variables on perceived relevance, manipulation, and content quality, modeled in latent space ($z$). See Table~\ref{tab:perception-cutpoints} for cut points associated with each perception and to assess effect sizes. Reference categories: gender = ``Man,'' age = ``25-34,'' education = ``Bachelor's degree.''}
    \label{fig:perceptions-demographics}
    \Description{Three subplots, one for each perception studied, showing how the demographic variables age, gender, and highest educational attainment influenced participants' ratings in the task.}
\end{figure*}

In contrast, participants' demographics influenced their perceptions on \popular{}. Figure~\ref{fig:perceptions-demographics} shows the effects of each demographic variable on perceived relevance, manipulation, and content quality, with each perception modeled as a latent variable $z$. This latent variable maps onto the Likert scale using the cut points estimated in Table~\ref{tab:perception-cutpoints}, and the distances between thresholds aid interpretation. For example, a $0.2$ increase in $z$ from $0.2$ to $0.4$ shifts the Likert rating from 2 to 3 for relevance, whereas the same increase starting at $0.0$ does not cross the $0.292$ threshold and therefore does not change the observed rating.

\subsubsection{Perceived Relevance}

Participants aged 18--24 and 35--44 perceived content as slightly more relevant than those aged 25--34, with $z$ increases of \separatehdi{0.183}{0.028}{0.347} and \separatehdi{0.131}{0.044}{0.223}, respectively. Women rated posts as less relevant \separatehdi{-0.151}{-0.224}{-0.079} than men. Master's and terminal degree holders perceived content as more relevant: \separatehdi{0.496}{0.369}{0.615} and \separatehdi{0.374}{0.155}{0.573}, respectively. Although statistically significant, these shifts are small relative to the average distance between relevance Likert thresholds ($\approx 1.0$). Overall, younger adults, women, and highly educated participants showed small but detectable differences in perceived relevance.

\subsubsection{Perceived Manipulation}

Participants aged 18--24 rated content as less manipulative \togetherhdi{-0.174}{-0.336}{-0.016} than the 25--34 reference group, while ages 35--44, 45--54, 55--64, and 65+ rated content as more manipulative, with increases in $z$ of \separatehdi{0.104}{0.013}{0.195}, \separatehdi{0.187}{0.090}{0.281}, \separatehdi{0.509}{0.378}{0.636}, and \separatehdi{0.282}{0.087}{0.498}, respectively. Women perceived posts as less manipulative \togetherhdi{-0.093}{-0.166}{-0.023} when compared to men. Compared to bachelor's degree holders, participants with less than high school, high school, some college, or associate's degrees perceived content as less manipulative (around $-0.2$ in $z$), except for the smallest group (less than high school), which had \separatehdi{-0.659}{-1.048}{-0.269}. Terminal degree holders \counts{2.91}{17} also rated content as less manipulative \togetherhdi{-0.223}{-0.431}{-0.007}. Overall, older participants and those with lower educational attainment (excluding terminal degrees) perceived posts as slightly more manipulative.

\subsubsection{Perceived Content Quality}

Participants aged 35--44, 45--54, and 55--64 rated posts as slightly lower quality than participants aged 25--34, with decreases of approximately $-0.14$ for the first two groups and \separatehdi{-0.265}{-0.393}{-0.138} for ages 55--64. Women also rated content lower \togetherhdi{-0.109}{-0.182}{-0.036} when compared to men. Education effects were minimal except for master's and terminal degree holders, who perceived content as higher quality: \separatehdi{0.359}{0.239}{0.482} and \separatehdi{0.242}{0.032}{0.446}, respectively. Overall, middle-aged participants and women perceived content as slightly lower quality, while highly educated participants perceived it as slightly higher quality.

Overall, demographic effects on perceptions were measurable but modest. As Table~\ref{tab:perception-cutpoints} shows, shifts rarely correspond to a full-step change on the Likert scale, given thresholds spaced by more than 0.9. Thus, demographic differences are statistically detectable but substantively minor.

\newcommand{\subredditonselectionfinding}{\textcolor{black}{\separatehdi{109.80\%}{50.08\%}{200.72\%}}}
\newcommand{\topiconselectionfinding}{\textcolor{black}{\separatehdi{40.49\%}{29.18\%}{54.19\%}}}

\subsection{RQ4: Influence of Content Preferences on Selection \& Perceptions}

\begin{figure*}
    \centering
    \includegraphics[width=0.75\linewidth]{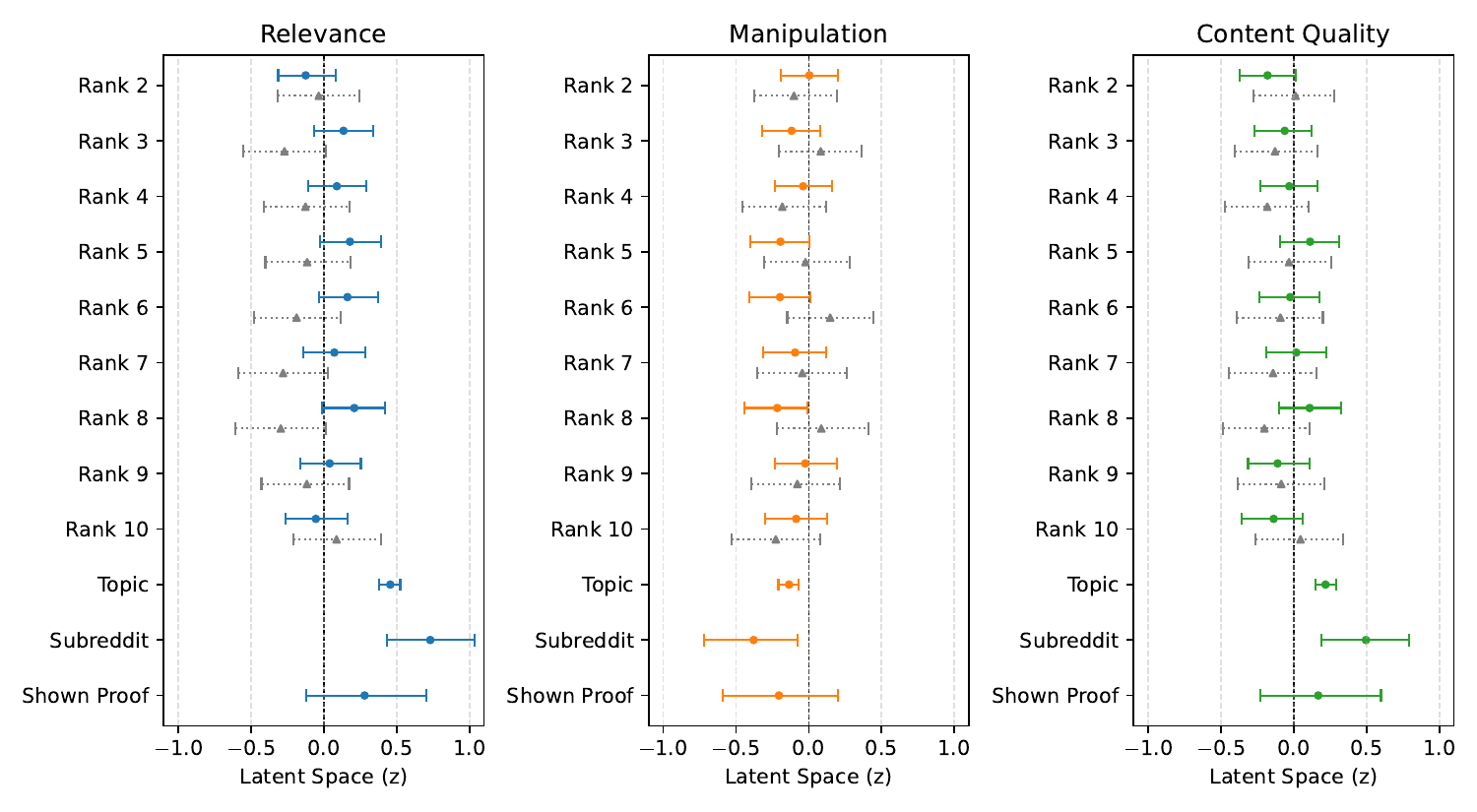}
    \caption{
        Effects of post rank and visibility of social proof on perceived relevance, trustworthiness, and content quality, modeled in a latent space ($z$). See Table~\ref{tab:perception-cutpoints} for the thresholds associated with each perception. Rank is treated categorically, with rank 1 as the reference. For rank, the blue circles denote the main effect of rank whereas the gray triangles denote the interaction effect with shown proof. The “Topic” and “Subreddit” labels indicate whether the participant mentioned the topic or subreddit during the pre-experiment questionnaire.
    }
    \label{fig:perceptions-main}
    \Description{Three subplots, one for each perception studied, showing how rank, social proof, and whether the post or subreddit was of interest influenced participants' ratings. The figure depicts both the main effect of social proof on each perception and its interaction with rank on the feed.}
\end{figure*}

The pre-experiment questionnaire asked participants to select topics they were interested in and to list at least five subreddits they regularly browsed. Using these responses, we examined whether participants’ content preferences influenced their selection decisions and their perceptions of posts on \popular{}.

Regarding selection, both topic and subreddit interest corresponded to substantial increases in the odds of a post being chosen. When a participant had indicated interest in the topic of a post (e.g., selecting “News \& Politics” in the questionnaire and then encountering a political post on \popular{}), the odds of selection increased by \topiconselectionfinding{}. Subreddit interest, although less frequent due to our corpus including only 32 subreddits, produced an even larger effect. Among the 585 participants, only 124 later encountered a subreddit they had listed in their pre-experiment questionnaire, and in these cases the odds of selection increased by \subredditonselectionfinding{}. These findings suggest that content preferences significantly shaped which posts participants chose to view.

Content preferences also influenced perceptions of posts. As shown at the bottom of Figure~\ref{fig:perceptions-main}, both topic and subreddit interest increased perceived relevance, which is intuitive since participants had already indicated interest in the subject matter. More precisely, topic interest increased perceived relevance by \separatehdi{0.418}{0.332}{0.496}, which corresponds to about a half-point increase on the Likert scale given the cut points in Table~\ref{tab:perception-cutpoints}, while subreddit interest produced an even stronger effect, increasing relevance by \separatehdi{0.716}{0.376}{1.077}. In terms of manipulation, topic interest slightly reduced perceived manipulativeness by \separatehdi{-0.182}{-0.268}{-0.098}, while subreddit interest yielded a larger reduction of \separatehdi{-0.425}{-0.797}{-0.070}. Finally, for perceived content quality, topic interest raised evaluations by \separatehdi{0.159}{0.076}{0.242}, and subreddit interest produced an even stronger increase of \separatehdi{0.510}{0.155}{0.840}.

Taken together, these results suggest that participants' prior interests in a topic or subreddit made posts appear more relevant and of higher quality while also seeming less manipulative. Although subreddit interest often produced larger effects than topic interest, the overlapping HDIs mean that these differences are not conclusive.

\begin{table*}[t]
    \centering
    \caption{Number and percentage of selections by rank across conditions (Both, Proof, and No Proof).}
    \begin{tabular}{lrrrrrr}
        \toprule
        & \multicolumn{2}{c}{\textbf{Both}} & \multicolumn{2}{c}{\textbf{Proof}} & \multicolumn{2}{c}{\textbf{No Proof}} \\
        \cmidrule(lr){2-3} \cmidrule(lr){4-5} \cmidrule(lr){6-7}
        \textbf{Rank} & \textbf{Num. Selected} & \textbf{\%} & \textbf{Num. Selected} & \textbf{\%} & \textbf{Num. Selected} & \textbf{\%} \\
        \midrule
        1  & 628 & 13.27\% & 322 & 13.98\% & 306 & 12.60\% \\
        2  & 510 & 10.78\% & 248 & 10.76\% & 262 & 10.79\% \\
        3  & 506 & 10.69\% & 236 & 10.24\% & 270 & 11.12\% \\
        4  & 545 & 11.51\% & 251 & 10.89\% & 294 & 12.10\% \\
        5  & 488 & 10.31\% & 245 & 10.63\% & 243 & 10.00\% \\
        6  & 456 & 9.63\%  & 226 & 9.81\%  & 230 & 9.47\%  \\
        7  & 391 & 8.26\%  & 176 & 7.64\%  & 215 & 8.85\%  \\
        8  & 397 & 8.39\%  & 199 & 8.64\%  & 198 & 8.15\%  \\
        9  & 386 & 8.16\%  & 192 & 8.33\%  & 194 & 7.99\%  \\
        10 & 426 & 9.00\%  & 209 & 9.07\%  & 217 & 8.93\%  \\
        \midrule
        \textbf{Total} & 4,733 & 100\% & 2,304 & 100\% & 2,429 & 100\% \\
        \bottomrule
    \end{tabular}
    \label{tab:selection-frequencies}
\end{table*}

% 	    No Proof	            Proof
%       Mean	2.5%	97.5%	Mean	2.5%	97.5%
% rank						
% 2	    -21.92	-36.01	-4.27	-23.75	-47.49	10.48
% 3	    -16.71	-31.84	1.50	-29.18	-52.09	2.50
% 4	    -6.81	-23.30	13.62	-22.93	-47.05	13.12
% 5	    -27.55	-41.30	-11.33	-26.27	-49.36	7.73
% 6	    -35.77	-47.84	-21.69	-34.76	-55.74	-6.22
% 7	    -40.11	-51.72	-26.75	-53.25	-68.37	-31.56
% 8	    -46.70	-56.78	-34.08	-44.04	-61.55	-17.99
% 9	    -47.62	-57.37	-34.82	-46.34	-63.11	-20.80
% 10	-39.46	-51.42	-26.20	-39.74	-60.07	-14.15

\newcommand{\ns}{\phantom{*}}

\begin{table}
    \centering
    \small
    \caption{Change in selection odds relative to rank~1 for both the no proof and proof conditions. These values are visualized in Figure~\ref{fig:selection-ranks}. Estimates are derived from the posterior distributions of the selection model described in Section~\ref{model:selection}. An asterisk (*) on the mean indicates that the 95\% high-density interval (HDI) does not include zero, indicating statistical significance. Ranks 2 and 5--10 in the ``no proof'' condition and ranks 6--10 in the ``proof'' condition had significantly lower selection odds compared to rank 1.}
    \begin{tabular}{l rr rr}
        \toprule
        & \multicolumn{2}{c}{\textbf{No Proof}} 
        & \multicolumn{2}{c}{\textbf{Proof}} \\
        \cmidrule(lr){2-3} \cmidrule(lr){4-5}
        \textbf{Rank} & \textbf{Mean} & \textbf{95\% HDI} & \textbf{Mean} & \textbf{95\% HDI} \\
        \midrule
        2  & $-21.9\%$*  & $[-36.0\%, -4.3\%]$   & $-23.8\%$\ns & $[-47.5\%, 10.5\%]$ \\
        3  & $-16.7\%$\ns & $[-31.8\%,  1.5\%]$   & $-29.2\%$\ns & $[-52.1\%,  2.5\%]$ \\
        4  & $-6.8\%$\ns  & $[-23.3\%, 13.6\%]$   & $-22.9\%$\ns & $[-47.1\%, 13.1\%]$ \\
        5  & $-27.6\%$*  & $[-41.3\%, -11.3\%]$  & $-26.3\%$\ns & $[-49.4\%,  7.7\%]$ \\
        6  & $-35.8\%$*  & $[-47.8\%, -21.7\%]$  & $-34.8\%$*   & $[-55.7\%, -6.2\%]$ \\
        7  & $-40.1\%$*  & $[-51.7\%, -26.8\%]$  & $-53.3\%$*   & $[-68.4\%, -31.6\%]$ \\
        8  & $-46.7\%$*  & $[-56.8\%, -34.1\%]$  & $-44.0\%$*   & $[-61.6\%, -18.0\%]$ \\
        9  & $-47.6\%$*  & $[-57.4\%, -34.8\%]$  & $-46.3\%$*   & $[-63.1\%, -20.8\%]$ \\
        10 & $-39.5\%$*  & $[-51.4\%, -26.2\%]$  & $-39.7\%$*   & $[-60.1\%, -14.2\%]$ \\
        \bottomrule
    \end{tabular}
    \label{tab:selection-ranks}
\end{table}

\begin{figure}
    \centering
    \includegraphics[width=\linewidth]{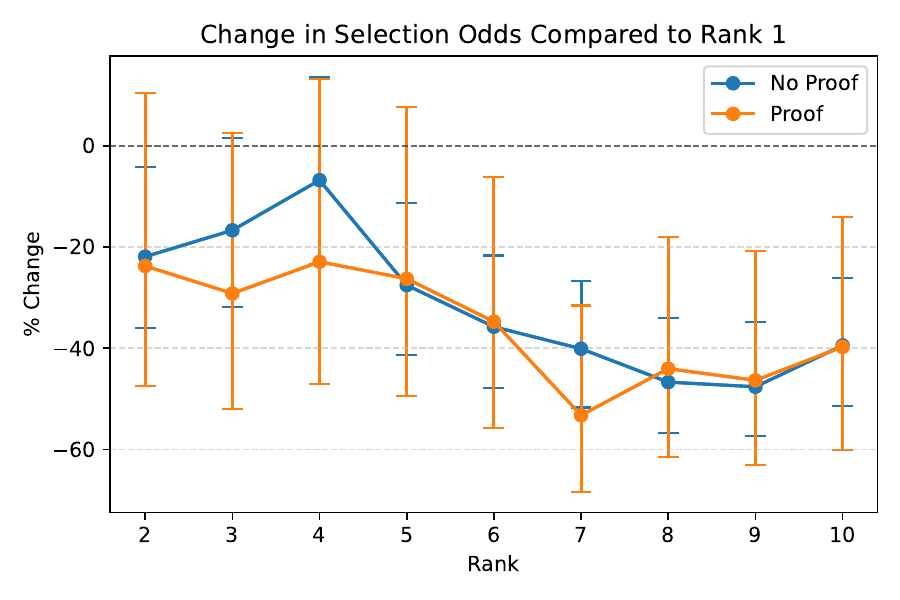}
    \caption{Change in selection odds relative to rank~1 for both the no-proof and proof conditions. Estimates come from the posterior distributions of the selection model described in Section~\ref{model:selection}. Error bars indicate 95\% high-density intervals (HDIs); results are significant when the HDI lies entirely below zero. In the no-proof condition, ranks 6--10 had significantly lower selection odds than rank~1, while in the proof condition this was true for ranks 7--9. Overall, placing the \textit{same} post in the lower half of the feed corresponds to an approximate $40\%$ reduction in the odds of being selected compared to placement at the top of \popular{}, regardless of social proof.}
    \label{fig:selection-ranks}
    \Description{Figure showing how the odds of selecting a post vary by its rank on the r/popular feed and whether social proof is displayed. Posts ranked below 5 had significantly lower odds of selection according to the regression models.}
\end{figure}

\subsection{H5: Influence of Rank on Selection \& Perceptions}

Table~\ref{tab:selection-frequencies} reports how often posts were selected at each rank under both social proof conditions, along with the combined frequencies. Even from this descriptive view, higher-ranked posts appear more likely to be selected than those positioned lower on \popular{}. But do these patterns hold in our regression models? And does rank influence perceived relevance, manipulation, or content quality?

Table~\ref{tab:selection-ranks} presents changes in selection odds relative to rank~1, our reference category, in both the proof and no proof conditions. Based on the high-density intervals, we found that in the no proof condition, ranks 2 and 5--10 had significantly lower odds of selection compared to rank~1. When social proof was shown, ranks 6--10 had significantly lower odds than rank~1. Figure~\ref{fig:selection-ranks} visualizes these changes as posts move down the \popular{} feed. Broadly, in both conditions, the lower half of the feed corresponded to an approximate $40\%$ reduction in selection odds compared to rank~1. An additional nuance is that rank~2 showed a significant reduction in the no proof condition, but this effect disappeared when social proof was present.

Because lower ranks were consistently less likely to be selected, we re-estimated the model using stronger priors for rank, treating it as a linear effect rather than categorical. This allowed us to test whether selection odds decreased consistently down the feed, and whether the interaction of rank with social proof was meaningful. Under this specification, the interaction effect was essentially nonexistent: with social proof, each decrease in rank corresponded to a \separatehdi{-6.62\%}{-8.20\%}{-5.03\%} change in selection odds, compared to \separatehdi{-6.62\%}{-8.11\%}{-5.00\%} without social proof. Given this and the overlapping HDIs in Table~\ref{tab:selection-ranks}, there is insufficient evidence that social proof amplifies rank-based inequality in engagement.

Finally, we examined whether perceptions were influenced by rank, and whether social proof intensifies these effects. As shown in Figure~\ref{fig:perceptions-main}, rank had no statistically significant effect on perceived relevance, manipulation, or content quality under either condition.

\subsection{H6: Influence of Social Proof on Selection \& Perceptions}

In addition to manipulating the ranks of each post, we also manipulated whether social proof was displayed on the feed. Social proof in this context encompassed two signals: the number of comments and the number of upvotes attached to each post. Accordingly, we examine not only whether the presence of social proof influenced selection behaviors and perceptions, but also whether the \textit{degree} of social proof (i.e., the magnitude of comments and upvotes) had any effect on participant choices or perceptions of posts on \popular{}.

% Selection (log-odds)
% shown_proof	                 0.131	-0.260	0.488
% shown_proof:num_comments_1000	-0.018	-0.098	0.055	
% shown_proof:num_upvotes_1000	 0.003	-0.005	0.012

From the selection model, we found no statistically significant effect of showing proof on the feed. In other words, the presence of social proof did not uniformly increase the odds of selection compared to a no-proof feed: \separatehdi{14.00\%}{-22.89\%}{62.91\%}. This aligns with the null results reported in the previous section and the lack of significant differences seen in Table~\ref{tab:selection-ranks}. We also found no statistically significant effect for the degree of social proof. Specifically, every 1,000 comments (the normalizing factor we used) corresponded to a \separatehdi{-1.78\%}{-9.06\%}{5.97\%} decrease in selection odds, while every 1,000 upvotes corresponded to a \separatehdi{0.30\%}{-0.60\%}{1.21\%} increase in odds. Given the wide HDIs, the model did not have sufficient evidence to conclude that either comments or upvotes systematically improved or worsened a post's chances of being selected.

% Perceptions: shown_proof main effect
% shown_proof	 0.280	-0.124	0.664
% shown_proof	-0.205	-0.579	0.181
% shown_proof	 0.168	-0.244	0.554

% Perceptions: degree of social proof
% shown_proof:num_upvotes_1000	-0.005	-0.015	0.005
% shown_proof:num_comments_1000	-0.059	-0.145	0.024
% shown_proof:num_upvotes_1000	 0.009	 0.000	0.019	
% shown_proof:num_comments_1000	-0.012	-0.088	0.067
% shown_proof:num_upvotes_1000	 0.002	-0.008	0.012
% shown_proof:num_comments_1000	-0.056	-0.136	0.026

Similarly, whether participants were shown social proof did not significantly change their perceptions of posts across the three measures we examined. Figure~\ref{fig:perceptions-main} shows that the main effects for social proof intersect with zero, indicating that the model could not distinguish a positive or negative influence. The degree of social proof also had no measurable effect on perceptions: coefficients for both comments and upvotes were essentially zero. For instance, every 1,000 upvotes shifted perceived relevance by only \separatehdi{-0.005}{-0.015}{0.005}. Results were nearly identical for other perception outcomes and for comments as well.

\section{Discussion} \label{sec:discussion}

Trending feeds are a central feature of social media platforms, drawing millions of users to browse what is presented as the platform's most relevant or popular content. They operate as shared spaces where collective culture is negotiated and reshaped~\cite{trending-is-trending}. Motivated by this, we set out to examine how two core mechanisms of these feeds---rank and social proof---shape users' behaviors and perceptions. Our study represents the first large-scale ($n=585$) randomized user experiments to investigate user experiences with trending feeds directly. Unlike prior observational studies~\cite{chan-popular-audit, chan-community-resilience, increased-attention-github, coordinated-campaigns}, our design allowed us to fully control post rankings and ensure all participants viewed the same set of posts. Specifically, our study design allowed us to examine ``What happens when the same post is shown to users at different ranks on the feed?'' This approach provides clear causal evidence about how rank influences engagement on highly visible feeds.

Our findings have implications for multiple stakeholders, including moderators, platform administrators, end users, and policymakers. Because trending feeds often function as digital ``town centers''~\cite{trending-is-trending}, it is crucial to understand whether they amplify biases in what topics gain attention, how users perceive content, and who benefits from visibility. Our results highlight the importance of monitoring these dynamics, particularly given broader concerns around misinformation, filter bubbles, and unequal treatment of vulnerable communities online~\cite{vendetta-black-creators}. 

Next, we discuss how our results inform ongoing debates about algorithmic curation, and how they lay the groundwork for future research on the design of trending feeds.

\subsection{Awareness of Social Media Algorithms}

Prior work has consistently shown that social media algorithms are often invisible to users~\cite{algorithm-awareness-chi-ea, alvarado-algorithmic-experience}. For example, \citet{invisible-news-feed} found that more than half of their 40 participants were unaware that Facebook's News Feed was algorithmically curated. Similarly, \citet{resnick-youtube-scrubbing} reported that nearly half of surveyed YouTube users ($44\%$) did not know the ``Not interested'' button was designed to influence recommendations. Even when platforms provide affordances for user control, these features frequently go unnoticed.

This lack of awareness, combined with the growing dominance of algorithmic curation, raises concerns about how content is surfaced online. Short-form video platforms such as TikTok, Instagram Reels, and YouTube Shorts now rely almost entirely on opaque algorithms to determine what users see, intensifying these concerns. Creators, in turn, have voiced increasing frustration with these black-box systems, which dictate visibility while offering little transparency or recourse~\cite{creator-friendly-youtube, youtube-algorithm-appeasement, tiktok-visibility-trap}.

Our study adds to this discussion by showing that while participants rarely self-reported rank as an explicit factor influencing their choices, our findings show that post selection on \popular{} are strongly influenced by a post's rank on the feed. In other words, users may not consciously recognize rank as influential, but their behavior reflects a systematic, implicit bias toward higher-ranked posts. This effect parallels findings from research on search engines~\cite{joachims-clicks}, yet differs in that trending feeds surface diverse posts without a user-supplied query.

These findings raise important questions for future work. Can greater transparency or explanations of ranking algorithms meaningfully empower users? Do such interventions encourage more responsible content consumption, or enable users to self-correct biases affecting their engagement and perceptions of posts? Early work on algorithmic explanations in Facebook's News Feed suggests limited benefits, in part because explanations are too generic or disconnected from individual users~\cite{news-feed-fyi, rader-explanations-facebook, tiktok-explanations}. More research is needed on effective, user-centered explanations that both inform and empower. Overall, our results underscore how strongly platforms can shape user behavior through ranking---even when users are unaware of it---highlighting the need for tools and policies that give users greater awareness and agency in algorithmically curated environments.

\subsection{Demographic Variance in Perceived Trust}

We observed a surprising pattern related to perceived trustworthiness and manipulation: older participants were consistently more skeptical of content on \popular{}. Specifically, those aged 35--44, 45--54, 55--64, and 65+ all rated posts as more manipulative compared to our reference group of 25--34 year olds. The item they endorsed at higher rates was: ``This post seems exaggerated or misleading to attract attention.'' It is difficult to determine whether this skepticism reflects accurate detection of attention-grabbing content or a broader distrust of online information among older participants. Further analysis would be required to establish whether these judgments are well-founded or indicative of a generalized suspicion of internet content.

We also found education-based differences. Participants with less than a bachelor's degree (i.e., less than high school, high school, or associate degree holders) rated posts on \popular{} as significantly less manipulative than participants with a bachelor's degree. Future work could examine whether these differences in baseline perceptions map onto more or less effective strategies for navigating online content. Without richer qualitative data (e.g., interviews) or experimental manipulations of post features, however, such conclusions remain speculative.

Interestingly, our findings diverge from \citet{tweeting-believing}, who found no significant demographic effects of age or gender on perceived credibility in the context of individual posts on Twitter (now X). A key distinction is that their study examined isolated posts on specific topics (politics, science, entertainment), whereas our participants evaluated posts within the broader context of a feed, where expectations and heuristics may differ.

Overall, we found that older participants and those with higher education levels tended to perceive posts on \popular{} as more manipulative or less trustworthy. These results suggest fruitful directions for future work on demographic differences in trust judgments, the strategies various groups employ when evaluating social media content, and whether such strategies can be leveraged to help mitigate misinformation.

\subsection{Rank-Based Inequality \& Dissipating Engagement}

Consistent with prior observational work on \popular{}~\cite{chan-popular-audit}, our experiment shows that the top-ranked post consistently receives a disproportionate share of attention and engagement, regardless of its actual content. This concentration of attention raises a natural question: is such inequality desirable for trending feeds? Should platforms embrace this dynamic as a feature, or should they consider mechanisms to distribute engagement more evenly?

One possible approach would be to reduce the determinism of ranking by introducing randomization. For example, platforms might shuffle the top $n$ posts, or periodically reorder every $m$ posts in an infinite-scrolling feed. Such interventions could blunt the strong positional advantage of the very top post. They might also yield engagement signals---such as upvotes, shares, or comments—that are less biased by rank and more reflective of content quality in aggregate. However, we acknowledge that introducing randomization may not align with platform goals for consistency and predictability. Alternative strategies could therefore focus on ways to present content accurately while still reducing positional bias---for example, supplementing ranks with explanatory context, grouping posts by topic, or incorporating additional signals that highlight quality without overemphasizing position.

Without such interventions, ranking tends to create a self-rein\-forc\-ing cycle: the top post is advantaged by its position, receives more engagement, and further widens the gap between itself and lower-ranked posts. This ``rich-get-richer'' dynamic, often described as the Matthew effect~\cite{matthew-effect}, is not unique to social media feeds. In other domains, such as online music markets, \citet{watts-music-markets} showed that rankings combined with social influence produced both inequality and unpredictability. In their parallel-worlds experiment, groups exposed to ranked feeds of songs exhibited higher inequality (as measured by Gini coefficients), and song quality explained only part of the variation in listens once social influence via download counts and live rankings was introduced.

\subsection{Design Implications}

Our findings have implications for the design of algorithmic ranking and curation systems, and more broadly, for social media platforms.

Since rank influences engagement implicitly---i.e., without the participant being aware---platforms should consider providing users with clearer explanations of how content is ranked in feeds. Prior work has highlighted both the opportunities and challenges of algorithmic explanations~\cite{tiktok-explanations, rader-explanations-facebook, alvarado-algorithmic-experience}. Building on this line of work, our results underscore the value of explanations that increase awareness of ranking signals. For example, platforms might surface post-level rationales such as ``this post appears because it received high engagement'' or ``this post appears because it matches your interests.'' Future research should test whether such explanations actually reduce the implicit biases of ranking systems, but continued experimentation in this area can help promote user awareness and agency.  

Second, designers might explore lightweight warnings or disclosures that acknowledge the role of engagement metrics (e.g., upvotes, clicks, retention) in shaping trending feeds. Making these mechanisms explicit could help mitigate overreliance on algorithmic signals. At the same time, explanations must balance technical accuracy with clarity and usability: if they are too complex, they risk being ignored; if too vague, they may mislead. Addressing these trade-offs is an important direction for both design and future research.  

Third, designers and researchers can play a critical role in developing alternative ranking strategies that go beyond purely engage\-ment-based feed algorithms. For instance, \citet{alexandria-reranking} explored incorporating LLM-generated values, such as honesty, in X's feed, while \citet{embedding-democratic-values} demonstrated how algorithms could be tuned to reduce partisan animosity by downranking certain posts. Such approaches suggest that feeds could prioritize qualitative values alongside (or even in place of) engagement. Making these options visible to users might also raise awareness of algorithmic curation and empower them to tailor their feeds to align with their own preferences. In some cases, customization could even foster collective practices (e.g., sharing how others configure their feeds), as seen with the collaborative use of Twitter blocklists~\cite{jhaver-blocklists}. Of course, these design interventions may conflict with platforms' commercial incentives to maximize engagement, but they highlight possible directions for rethinking ranking systems in ways that balance business goals with user empowerment.

Fourth, researchers and third-party developers can take the initiative by building tools that enable greater feed customization and experiment with alternative ranking strategies. While such tools and design interventions may initially reach only a subset of users, they can serve as valuable testbeds for evaluating the effectiveness of different approaches and building an evidence base to encourage adoption by major platforms. For example, although not a third-party feed, \citet{wang2024lower} provide a strong methodological precedent: their within-subject experiment compared chronological and algorithmically curated feeds on X/Twitter, showing that while feed design altered user behaviors, perceptions of the platform itself were more resilient across conditions. Similar experiments could help researchers and developers empirically demonstrate the potential benefits of alternative ranking systems in practice.

Overall, improving the visibility of ranking algorithms---through explanations, warnings, or other forms of trans\-parency---has the potential to empower users and give them greater control over how they interpret and engage with social media content.

\subsection{Directions for Future Researchers to Explore}

Our experiment did not find sufficient evidence that rank within the top 10 of \popular{} or the presence and degree of social proof significantly altered perceptions of relevance, trustworthiness, or content quality. Rather than suggesting these factors never matter, our results indicate that their influence may be conditional on the design of the feed and the context in which users encounter posts. Below, we outline several possible explanations and highlight how they can guide future research on rank and user perceptions.

First, the span of ranks we examined may not have been wide enough to elicit perceptual differences. Prior work on \popular{} shows that engagement differences among posts in the top 10 are relatively small, whereas sharper drop-offs occur much lower in the feed~\cite{chan-popular-audit}. Participants in our study spent a median of 55 seconds browsing ten posts ($\mu=1m\ 1s$, $\sigma=35s$). With such short feeds, the difference between rank 1 and rank 10 may not have been especially salient. Stronger rank effects may emerge in longer feeds (e.g., rank 1 vs.\ rank 50), where users must scroll further and rank carries greater informational weight. While experimentally controlling longer feeds can be resource-intensive, longitudinal methods---similar to \citet{reddit-usage-data}---offer a promising way to capture these effects in the wild.

Second, the engagement signals we varied---i.e., switching social proof on or off---may have had limited discriminating power in our corpus. Most posts already had tens of thousands of upvotes and hundreds of comments. Under such conditions, toggling visibility may not meaningfully shift perceptions. This pattern aligns with \citet{more-stars-reviews}, who found that reputation signals such as ratings or reviews matter less when items are already clearly differentiated. Our findings thus suggest that social proof may be most influential when engagement levels are more ambiguous or contested.

Third, topic diversity on \popular{} may reduce reliance on rank or social proof. Our corpus of 50 posts spanned 36 unique communities, with no subreddit contributing more than three posts. In contrast, search engine result page studies~\cite{serps-conflicting-science, joachims-clicks, biased-search-debated-topics} examine homogeneous result sets (e.g., multiple abstracts for the same query), where users must depend more on algorithmic cues to choose. In our case, heterogeneous topics likely enabled participants to evaluate each post on its own merits rather than relying on rank or engagement metrics.

Fourth, when participants rate posts, a pop-up window appears to collect their perceptions for relevance, trustworthiness, and content quality. Thus, the feed is taken over by the pop-up and they only see the post they are supposed to rate (see~\autoref{fig:popup}). It may be possible that participants are decontextualized from ranking's influence when the pop-up appears. Future work should examine whether having participants rate content directly on the feed, as opposed to have a pop-up window appear, leads to different results. Alternatively, participants could also be asked to select the top three posts in terms of quality or relevance, akin to the task in our selection phase.

Taken together, these considerations suggest that rank and social proof are not universally powerful cues but operate as conditional signals, especially when users must discriminate among otherwise similar options. Future work can test these dynamics by extending feeds to longer lengths, targeting more homogeneous topic contexts (e.g., r/news), or employing longitudinal designs that periodically prompt users for perceptions. Our findings therefore refine understanding of when and how feed signals shape user judgments: when posts are already high-quality and diverse, such signals may play a smaller role, whereas in more competitive or ambiguous contexts they may be decisive.

\subsection{Limitations \& Future Work}

Our experiment asked participants to interact with static image captures (i.e., snapshots) of the \popular{} feed, where they indicated which posts they would engage with and rated each post's relevance, trustworthiness, and content quality. While this approach allowed us to precisely manipulate rank and social proof, while keeping the same set of posts, it also introduced limitations. For example, image galleries only displayed the first image, and video posts were reduced to a single still frame. Participants therefore lacked some information they would have when browsing \popular{} directly. That said, \citet{reddit-usage-data} found that users rarely click through before voting ($73\%$), suggesting screenshots capture much of the relevant decision context. Moreover, only one out of 585 participants explicitly mentioned missing information in the exit questionnaire, indicating that this issue was not widespread.

Another limitation concerns timing. The earliest snapshot was collected on July 15, 2025, and the last participant completed the study on August 22, 2025. While this five-week lag raises the possibility that content aged out of relevance, no participants reported this as a concern in the exit questionnaire. In fact, when asked how similar the posts were to what they normally see in their feeds, 475 participants ($81.2\%$) rated the content a 4 or 5 on a five-point scale, reinforcing that the snapshots felt realistic.

The scope of our corpus also constrains generalizability. We used five snapshots (50 posts) which included 36 subreddits, with no subreddit represented more than three times. While this provided sufficient variation for statistical analysis, it limited our ability to capture the full breadth of \popular{}. Notably, 17 of the posts were political, many referencing the new U.S. presidential administration. Several participants reported political fatigue in the exit questionnaire, suggesting that the political salience of our sample may have influenced results.

Future work can address these limitations by expanding the number of posts shown or by focusing on topic-specific feeds and communities. It may also be valuable to examine additional feed signals, such as awards or post flairs, to see how they shape user behavior. Another promising direction is longitudinal research that passively observes user behavior through browser extensions or in-situ prompts, allowing perceptions to be captured in more naturalistic settings. As popular and trending feeds continue to concentrate attention and shape online discourse, it is crucial to understand how their embedded signals guide engagement and influence cultural conversation.

\section{Conclusion}

In this paper, we conducted a large-scale ($n=585$) randomized experiment in which participants engaged with static snapshots of Reddit’s trending feed, \popular{}. We examined how users chose which posts to engage with and how they perceived content across three dimensions: relevance, trustworthiness, and quality. Our analysis shows that algorithmic rank plays a role in directing attention, even though participants rarely reported rank as a conscious factor in their decisions. At the same time, perceptions of content did not significantly vary by rank within the top 10 posts, suggesting that ranking may shape visibility more than evaluation---or that the top 10 was not enough to elicit perceptual differences among participants. We also expected post-level engagement metrics (i.e., social proof) to amplify inequality by reinforcing rank effects, but found little evidence of this, indicating that these metrics were not especially useful discriminators for participants.  

Taken together, these findings contribute to ongoing discussions about social media feeds and the opaque algorithms that drive them. As platforms increasingly shape what information users encounter, understanding how ranking and other feed signals guide attention and perception remains essential. Our study underscores both the subtle influence of algorithmic ordering and the limits of engagement metrics, pointing toward the need for more transparent and accountable feed design as these systems continue to mediate online content consumption across platforms.

\bibliographystyle{ACM-Reference-Format}
\bibliography{bib}

\newpage

\appendix

\section{Demographic Information} \label{sec:demographics}

% Man                  322
% Woman                250
% Non-binary            10
% Prefer not to say      3

% Gender table
\begin{table}[h]
    \centering
    \caption{Gender distribution}
    \begin{tabular}{lrr}
        \toprule
        \textbf{Gender} & $n$ & \% \\
        \midrule
        Man                  & 322 & 55.02\% \\
        Woman                & 250 & 42.73\% \\
        Non-binary           &  10 &  1.71\% \\
        Prefer not to say    &   3 &  0.51\% \\
        \midrule
        \textbf{Total}       & 585 & 100.00\% \\
        \bottomrule
    \end{tabular}
\end{table}

% Less than high school                               5
% High school graduate or equivalent (e.g., GED)     68
% Some college, no degree                           116
% Associate degree (e.g., AA, AS)                    80
% Bachelor's degree (e.g., BA, BS)                  237
% Master's degree (e.g., MA, MS, MBA)                61
% Graduate degree (e.g., PhD, MD, JD)                17
% Prefer not to say                                   1

% Education table
\begin{table}[h]
    \centering
    \caption{Education distribution}
    \begin{tabular}{lrr}
        \toprule
        \textbf{Education} & $n$ & \% \\
        \midrule
        Less than high school                              &   5 &  0.85\% \\
        High school graduate or equivalent (e.g., GED)     &  68 & 11.62\% \\
        Some college, no degree                            & 116 & 19.83\% \\
        Associate degree (e.g., AA, AS)                    &  80 & 13.68\% \\
        Bachelor's degree (e.g., BA, BS)                   & 237 & 40.51\% \\
        Master's degree (e.g., MA, MS, MBA)                &  61 & 10.43\% \\
        Graduate degree (e.g., PhD, MD, JD)                &  17 &  2.91\% \\
        Prefer not to say                                  &   1 &  0.17\% \\
        \midrule
        \textbf{Total}                                     & 585 & 100.00\% \\
        \bottomrule
    \end{tabular}
\end{table}

% 18-24                 34
% 25-34                191
% 35-44                160
% 45-54                119
% 55-64                 60
% 65 or older           19
% Prefer not to say      2

% Age table
\begin{table}[h]
    \centering
    \caption{Age distribution}
    \begin{tabular}{lrr}
        \toprule
        \textbf{Age} & $n$ & \% \\
        \midrule
        18--24                 &  34 &  5.81\% \\
        25--34                 & 191 & 32.65\% \\
        35--44                 & 160 & 27.35\% \\
        45--54                 & 119 & 20.34\% \\
        55--64                 &  60 & 10.26\% \\
        65 or older            &  19 &  3.25\% \\
        Prefer not to say      &   2 &  0.34\% \\
        \midrule
        \textbf{Total}         & 585 & 100.00\% \\
        \bottomrule
    \end{tabular}
\end{table}

\section{Experimental Condition Distribution}

Refer to Table~\ref{tab:snapshots} for the number of participants in each experimental condition.

\begin{table*}[t]
    \caption{The number of participants who viewed each condition. Note that each participant views three feeds during the experiment. The rows indicate which set of posts they saw and the columns indicate how the posts were rotated (e.g., R1 indicates the posts were rotated down one position). The first number in each cell denotes the number of views in the social proof conditions whereas the latter is when there was no social proof shown. For example, ``11/26'' indicates that 11 participants saw the feed with social proof and 26 did not.}
    \centering
    \begin{tabular}{lrrrrrrrrrrr}
        \toprule
         & \textbf{R0} & \textbf{R1} & \textbf{R2} & \textbf{R3} & \textbf{R4} & \textbf{R5} & \textbf{R6} & \textbf{R7} & \textbf{R8} & \textbf{R9} & \textbf{Total} \\
        \midrule
        \textbf{Snapshot 1} & 11/26 & 18/22 & 15/18 & 18/13 & 23/24 & 20/11 & 24/16 & 12/17 & 20/17 & 14/17 & 175/181 \\
        \textbf{Snapshot 2} & 22/19 & 14/25 & 14/18 & 19/19 & 18/20 & 18/19 & 16/21 & 24/13 & 15/25 & 13/17 & 173/196 \\
        \textbf{Snapshot 3} & 14/20 & 18/11 & 19/31 & 11/15 & 17/21 & 11/13 & 18/25 & 20/16 & 15/15 & 22/15 & 165/182 \\
        \textbf{Snapshot 4} & 16/12 & 16/16 & 17/19 & 16/17 & 23/20 & 25/14 & 22/18 & 16/19 & 19/14 & 11/18 & 181/167 \\
        \textbf{Snapshot 5} & 21/21 & 17/17 & 16/19 & 12/11 & 14/18 & 12/22 & 19/19 & 19/18 & 15/9 & 13/23 & 158/177 \\
        \bottomrule \\
    \end{tabular}
    \label{tab:snapshots}
\end{table*}

% Proof
%                                       0	1	2	3	4	5	6	7	8	9	total
% 749fc4f5-ce90-4a98-b5e7-39db62f1632b	11	18	15	18	23	20	24	12	20	14	175
% a602674f-7bc6-47e7-919b-bcde0dcf5f05	22	14	14	19	18	18	16	24	15	13	173
% 1f9d2527-6430-4c8e-87cf-37a3e14323be	14	18	19	11	17	11	18	20	15	22	165
% c9dfbf3e-8655-4059-a1ac-fdb73fd88764	16	16	17	16	23	25	22	16	19	11	181
% cd299601-0ec4-49ab-a410-64489a867dea	21	17	16	12	14	12	19	19	15	13	158
% Condition matrix without proof shown:

% No Proof
%                                       0	1	2	3	4	5	6	7	8	9	total
% 749fc4f5-ce90-4a98-b5e7-39db62f1632b	26	22	18	13	24	11	16	17	17	17	181
% a602674f-7bc6-47e7-919b-bcde0dcf5f05	19	25	18	19	20	19	21	13	25	17	196
% 1f9d2527-6430-4c8e-87cf-37a3e14323be	20	11	31	15	21	13	25	16	15	15	182
% c9dfbf3e-8655-4059-a1ac-fdb73fd88764	12	16	19	17	20	14	18	19	14	18	167
% cd299601-0ec4-49ab-a410-64489a867dea	21	17	19	11	18	22	19	18	9	23	177

\section{Feed Rotations Example} \label{sec:full-interface}

\begin{figure*}
    \centering
    \begin{subfigure}{0.46\textwidth}
        \centering
        \includegraphics[width=0.45\textwidth]{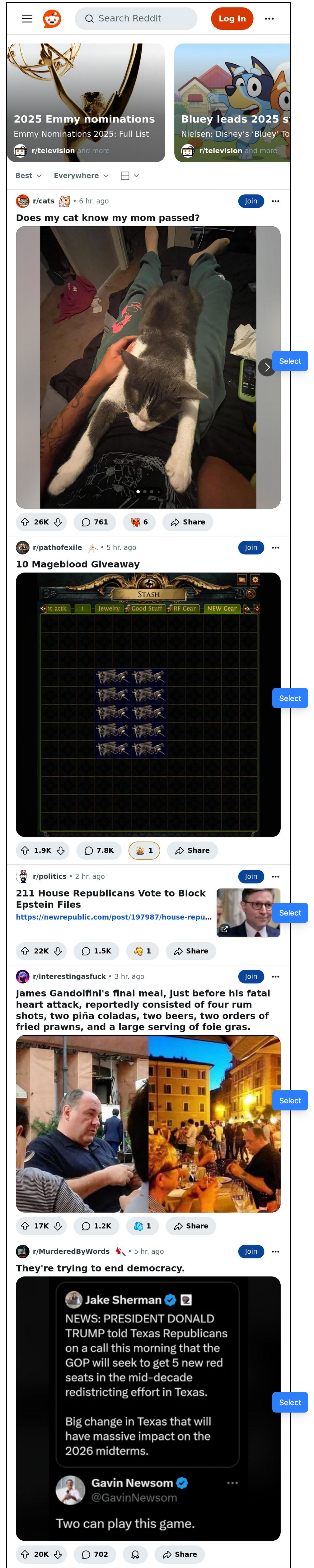}
        \caption{}
    \end{subfigure}
    \hfill
    \begin{subfigure}{0.46\textwidth}
        \centering
        \includegraphics[width=0.45\textwidth]{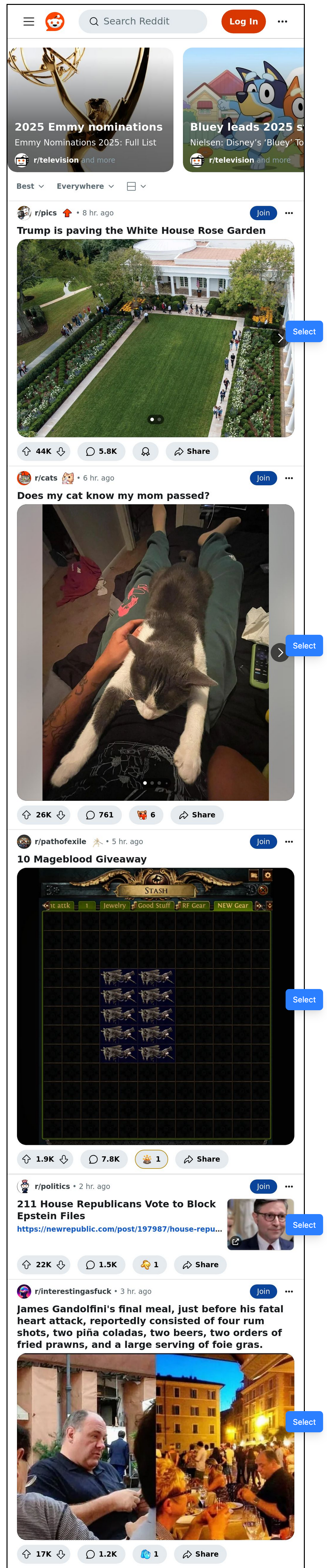}
        \caption{}
    \end{subfigure}
    \caption{
        Example feeds shown during the ``Selection'' phase (Section~\ref{sec:design}). Each post is accompanied by a ``Select'' button, which participants used to indicate engagement. Feed versions (a) and (b) illustrate our rotation procedure: to generate (b), we moved the tenth post to the top of the feed and shifted the remaining posts down by one position. In the subsequent ``Rating'' phase, the ``Select'' buttons were replaced with ``Rate'' buttons that opened a pop-up window for participants to evaluate the posts they had chosen.
    }
    \Description{
        Two screenshots of r/popular showing ten posts each. In the first screenshot (feed version a), posts are displayed in their original order. In the second screenshot (feed version b), the tenth post has been moved to the top, while the other nine posts have each shifted down one rank. This demonstrates the feed-rotation method we used to vary post positions while keeping the same set of ten posts, allowing us to isolate the effects of rank on user engagement and perceptions.
    }

    \label{fig:full-interface}
\end{figure*}

Figure~\ref{fig:full-interface} illustrates how posts are rotated on \popular{} and how each post has a respective ``Select'' button that participants used to indicate their engagement.

\clearpage

\section{Social Proof Visibility Example} \label{sec:social-proof}

\begin{figure}[h]
    \centering
    \begin{subfigure}{\linewidth}
        \centering
        \includegraphics[width=\textwidth]{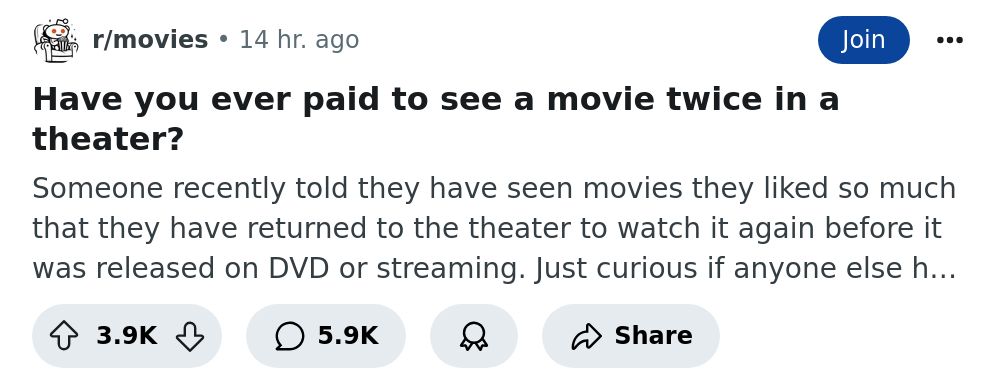}
        \caption{With social proof visible.}
    \end{subfigure}

    \vspace{2em}
    
    \begin{subfigure}{\linewidth}
        \centering
        \includegraphics[width=\textwidth]{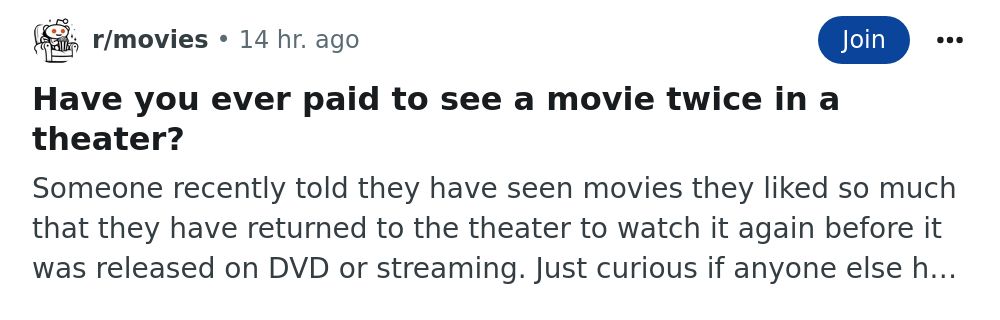}
        \caption{Without social proof visible.}
    \end{subfigure}
    \caption{
        Example demonstrating how we show and hide social proof---i.e., post-level engagement metrics like score and number of comments---for the same post. A feed being shown without social proof visible means all posts in the screenshot will not have engagement metrics shown.
    }
    \Description{
        Two screenshots of a single r/movies post. The first post has the engagement metrics showing below the post's content, i.e., the number of comments and upvotes as well as buttons for awards and sharing. The second screenshot is the same post, but without those engagement metrics.
    }

    \label{fig:social-proof}
\end{figure}

% Left: five posts, one rotation, social proof on
% Right: same five posts, one shifted down, social proof off

\end{document}